\documentclass[onecolumn]{emulateapj}

\usepackage{natbib}

\newcommand{\dfrac}[2]{{\displaystyle \frac{#1}{#2}}  }

\newcommand{\eqref}[1]{(\ref{#1})}

\def\lesssim{\mathrel{\hbox{\rlap{\hbox{\lower4pt\hbox{$\sim$}}}\hbox{$<$}}}}
\def\gtrsim{\mathrel{\hbox{\rlap{\hbox{\lower4pt\hbox{$\sim$}}}\hbox{$>$}}}}

\slugcomment{ApJ, Accepted}

\shorttitle{Gap Opening in Inviscid Disk}
\shortauthors{Muto, Suzuki and Inutsuka}

\begin{document}

\title{Two-Dimensional Study of the Propagation of Planetary Wake and
the Indication to Gap Opening in an Inviscid Protoplanetary Disk}

\author{Takayuki Muto\altaffilmark{1}}
\affil{Department of Earth and Planetary Sciences, 
Tokyo Institute of Technology, \\
2-12-1 Oh-okayama, Meguro-ku, Tokyo, 152-8551, Japan}

\author{Takeru K. Suzuki and Shu-ichiro Inutsuka}
\affil{Department of Physics, Nagoya University, \\
Furo-cho, Chikusa-ku, Nagoya, 464-8602, Japan}

\email{muto@geo.titech.ac.jp}
\altaffiltext{1}{JSPS Research Fellow}

\begin{abstract}
We analyze the physical processes of gap formation in an inviscid 
 protoplanetary disk with an embedded protoplanet using two-dimensional
 local shearing-sheet model.  
 Spiral density wave launched by the planet shocks and the angular
 momentum carried by the wave is transferred to the background flow.
 The exchange of the angular momentum can affect the mass flux in the
 vicinity of the planet to form an underdense region, 
or gap, around the planetary orbit.   
 We first perform weakly non-linear analyses to
 show that the specific vorticity formed by shock dissipation of density
 wave can be a source of mass flux in the vicinity of the planet, and
 that the gap can be opened even for low-mass planets
unless the migration of the planet is substantial. 
 We then perform high resolution numerical simulations 
 to check analytic consideration.
 By comparing the gap opening timescale and type I migration timescale, 
 we propose a criterion for the formation of underdense region around
 the planetary orbit that is qualitatively different
 from previous studies.
 The minimum mass required for the planet to form a dip is twice as
 small as previous studies if we incorporate the standard values of type
 I migration timescale, but it can be much smaller if there is a
 location in the disk where type I migration is halted.  
\end{abstract}


\section{Introduction} 
\label{sec:intro}

Disk-planet interaction is one of the important topics in the planet 
formation theory.  A low mass planet embedded in a disk excites the
density wave (Goldreich and Tremaine 1979), and the backreaction from
the density wave causes the planet to migrate in the disk (e.g., Ward
1986, Tanaka et al. 2002).  The excitation of density wave 
at Lindblad resonances can be
understood by linear analyses, although it has recently been pointed out
that non-linear effects are also important at corotation resonances
(Paardekooper and Papaloizou 2009).  
For a high mass planet, the interaction 
between the planet and the disk becomes nonlinear, and the gap opens
around the planetary orbit
(Lin and Papaloizou 1986ab, Ward and Hourigan 1989, 
Rafikov 2002b, Crida et al. 2006). 

Gap formation around the planet is important both theoretically and
observationally.  
From the theoretical point of view, gap formation determines the regime
of planetary migration.  If there is no gap opening, planetary migration
is in the regime called ``type I'', where the interaction between the
planet and the spiral density wave is important 
(Goldreich and Tremaine 1979, Ward 1986, 1997, Tanaka et al. 2002).  
Type I planetary
migration timescale is considered to be faster than disk dispersal
timescale, which poses a serious problem in the theory of planet
formation.  If the gap opens around the planet, the migration is in the
regime called ``type II'', where the planet migrates as the disk
accretion onto the central star occurs
(Lin and Papaloizou 1986).
The timescale of type II migration,
which is of the order of viscous timescale, 
is generally longer than type I migration, and there may be a
possibility for the planets to survive in the disk.  
From observational point of view, the gap around the planet may be able
to be observed by direct imaging.  Recent
progress of disk observation by direct imaging has reached the stage
that it is possible to compare the numerical simulations and observation
directly (Mayama et al. 2009).  
Moreover, the dynamical interaction between the circumstellar dust or
gas and the planet can be used to estimate the mass of a low-mass object
embedded in the disk (Kalas et al. 2008).  If the gap in the disk can
extend to the disk scale, the gap structure can be a very
good indicator of the existence of the planet.

Conventionally, gap opening is understood as a balance between the
torque exerted by the planet and the viscous torque (Lin and Papaloizou
1986b).  
In addition to the viscous torque, Crida et al. (2006) has shown
that ``pressure torque'' also acts to balance the planetary torque.
Gap opening processes are mainly
investigated using one-dimensional
model.  Rafikov (2002b) calculate the evolution of disk surface density
in the vicinity of the planet using a classical 
one-dimensional model (Lynden-Bell and Pringle 1974, Pringle 1981).  
He has shown that in the vicinity of
the low-mass planet, the torque exerted by the planet is carried away by
density wave
%%%%%%%%%%%%%%%%%%%%%%%%%%%%%
\footnote{
Note that Crida et al. (2006) explain this in terms of the balance
between the pressure torque and the planetary torque.
}
%%%%%%%%%%%%%%%%%%%%%%%%%%%%%
, and the gap is not opened until the wave shocks to deposit
the angular momentum to the mean flow.  
Considering the pressure effects as well as viscous effects, 
Crida et al. (2006) have derived the gap
opening criterion generalized for arbitrary values of kinematic
viscosity
using two-dimensional numerical simulation 
 (equation (15) of their paper).  Using the criterion of Crida
et al. (2006), the gap-opening mass for an inviscid disk reads 
\begin{equation}
 \dfrac{M_{\rm p}}{M_{\ast}} \gtrsim 1.6 \times 10^{-4}
 \left( \dfrac{H/r}{0.05} \right)^3,
 \label{cond_crida}
\end{equation}
where $M_{\rm p}$ is the mass of the planet, $M_{\ast}$ is the mass of
the central star, $H$ is the scale height of the disk, and $r$ is the
orbital semi-major axis of the planet.  Therefore, planets more massive
than Saturn can open up the gap in the disk.

However, recently, Li et
al. (2009) have performed high resolution numerical simulations 
and showed that a 
partial gap is formed in the vicinity of the planet even if the mass of
the plant is smaller than the mass limit given by equation
\eqref{cond_crida} in case the disk viscosity is very low.  This
motivates us to investigate the physical processes of gap opening in an
inviscid disk in detail.   

In this paper, we show that low mass planets can potentially open up a
partial gap.  
We investigate the processes by means of analytical models taking
into account weak non-linearity.  
We also perform numerical simulations to
look at to what extent numerical calculations and analytic studies
agree, and then finally we suggest the criterion for the gap opening in
an inviscid disk.  For analytic studies, we study the propagation of the
spiral density wave and subsequent shock formation.  We then investigate
the mass flux in the vicinity of the planet by means of second-order
perturbation theory.  We note that the second-order perturbation is
necessary to study the mass flux since it is essentially a second-order
quantity. 

The plan of this paper is as follows.  
In Section \ref{sec:basiceq}, we describe the basic equations.  In
Section \ref{sec:analytic}, we investigate the gap opening processes
using a second-order perturbation theory.  
We show that the shock dissipation of density wave and
the subsequent formation of specific vorticity 
can lead to the mass flux in the
vicinity of the planet, resulting in the gap formation.  We then show
the results of numerical simulations in Section \ref{sec:numerical}.  In
Section \ref{sec:gapform}, we discuss the condition for gap opening in
an inviscid disk.  We also discuss that commonly used one-dimensional
models of disk evolution may overlook gap opening processes in the
vicinity of the planet.  
Section \ref{sec:summary} is for summary.

\section{Basic Equations}
\label{sec:basiceq}

In this paper, we focus on the two-dimensional local shearing-sheet
analysis for simplicity.  
We set up a local Cartesian coordinate system 
corotating with a planet.  We take the origin of the coordinate system
at the planet's location, and the $x$- and $y$-axes are the radial and
the azimuthal direction, respectively.  We use isothermal ideal
hydrodynamic equations
\begin{equation}
 \dfrac{\partial \Sigma}{\partial t} + 
  \nabla \cdot (\Sigma \mathbf{v}) = 0
  \label{EoC_full}
\end{equation}
\begin{equation}
 \dfrac{\partial \mathbf{v}}{\partial t} + 
  \mathbf{v} \cdot \nabla \mathbf{v} = 
  -\dfrac{c^2}{\Sigma} \nabla{\Sigma} 
  - 2 \Omega_p \mathbf{e}_z \times \mathbf{v} 
  + 3 \Omega_p^2 x - \nabla \psi_{\rm p}
  \label{EoM_full}
\end{equation}
where we have assumed that the gas is rotating at the Kepler velocity.  
Notations are as follows: $\Sigma$ is surface density, 
$\mathbf{v}$ is velocity
field, $c$ is sound speed, $\Omega_{\rm p}$ is the angular velocity of
the planet, and $\psi_{\rm p}$ is the gravitational potential of the
planet.  For $\psi_{\rm p}$, we assume the form
\begin{equation}
 \psi_{\rm p} = 
  \dfrac{GM_{\rm p}}{\left( x^2 + y^2 + \epsilon^2 \right)^{1/2}},
\end{equation} 
where $G$, $M_{\rm p}$, and $\epsilon$ are the gravitational constant, 
the mass of the planet, and the softening parameter, respectively.

Local shearing-sheet approximation is only an approximation of global
model, and it may not be appropriate for investigating the global
evolution of the disk structure.  However, local
shearing-sheet approximation and full global model share many essential
physics in common.  Excitation and the propagation of
density wave can be understood using local approximation.  We also show
later that one-dimensional disk evolution model constructed from global
model and local model are very similar.  
Local approximation also has the advantages that analyses
are greatly simplified and high-resolution calculations become
possible.  

Here, we 
comment on the two-dimensional approximation.  It is known
that three-dimensional processes are important in disk-planet
interaction especially when considering the immediate vicinity of the
planet.  For example, 
numerical simulations by Paardekooper and Mellema (2006)
clearly show that disk structure around the planet is
three-dimensional.  However, for the study of density wave, which is the
structure away from the location of the planet, 
the density perturbation by the planet is nearly two-dimensional.   
In this paper, we shall investigate the physical properties of density
wave excited by the planet in detail, and we discuss the gap formation
processes that can be derived from the study of density wave.  
Therefore, 
we expect that essential physics can be captured by local
two-dimensional model.

\section{Analytic Study of Planetary Wake}
\label{sec:analytic}

In this section, we investigate the disk-planet interaction and
subsequent gap opening processes using analytic methods.  We consider a
weakly non-linear stage, where the mass of the planet is 
\begin{equation}
 \dfrac{GM_{\rm p}}{Hc^2} \lesssim 1, 
  \label{nonlin_criterion}
\end{equation}
which is approximately smaller than the Saturn mass in case of Minimum
Mass Solar Nebula (Hayashi et al. 1985) with $H/r\sim 0.05$
%%%%%%%%%%%%%%%%%%%%%%%%%%%%%%%%%%%
\footnote{
Equation \eqref{nonlin_criterion} is equivalent to the condition where
Hill's radius becomes comparable to the disk scale height.  Departure
from linear calculations can be observed at this mass range, see e.g.,
Miyoshi et al. (1999)
}
%%%%%%%%%%%%%%%%%%%%%%%%%%%%%%%%%%%
. 
We note that non-linearity of disk-planet interaction and gap opening is
strongly related.  From equation \eqref{nonlin_criterion}, the onset of
the non-linearity is given by 
\begin{equation}
 \dfrac{M_{\rm p}}{M_{\ast}} \gtrsim 
  1.25 \times 10^{-4} \left( \dfrac{H/r_{\rm p}}{0.05} \right)^3,
\end{equation}
where $H=c/\Omega_{\rm p}$ is the scale height of the disk.  This
non-linear criterion is analogous to equation \eqref{cond_crida}, 
which is the gap opening criterion given by Crida et al. (2006)

\begin{figure}
 \plotone{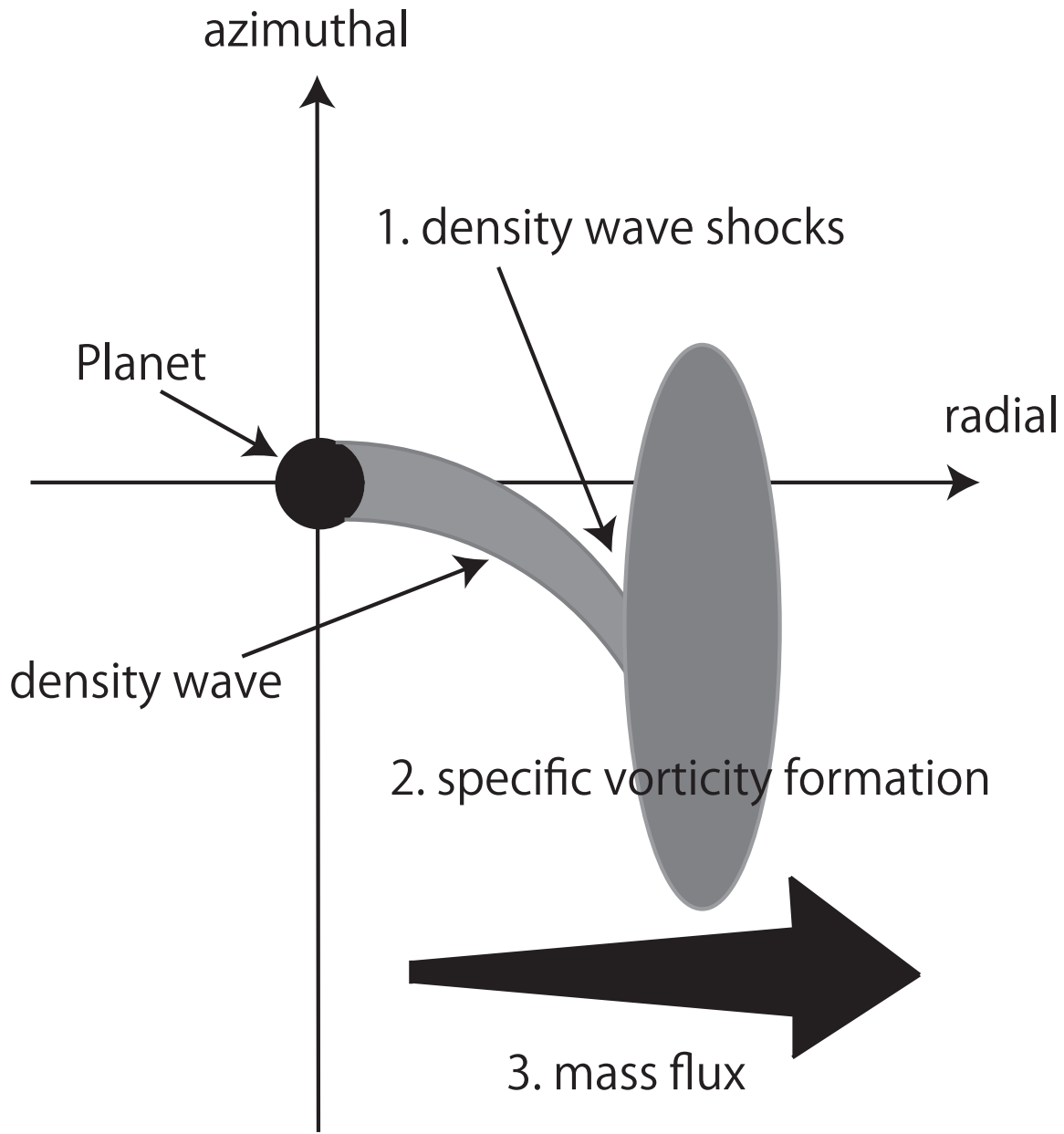}
 \caption{Overall picture of the physical processes involved in gap
 opening.  The density wave launched by the planet eventually shocks as
 it propagates in the radial direction.  Dissipation of the density wave
 then leads to the formation of the specific vorticity.  The specific
 vorticity leads to the mass flux around the planet, which results in
 the gap formation.  
 }
 \label{fig:gap_schematic}
\end{figure}

The overall picture of gap formation we suggest is as follows (see also
Figure \ref{fig:gap_schematic}).
\begin{enumerate}
 \item Density wave excited by the planet will shock at some location
       away from the planet.
 \item The shock formation leads to the formation of specific vorticity.
 \item The change of specific vorticity results in net radial mass flux.    
       This mass flux exists in the place closer to the planet, 
       even at the place where 
       the change of specific vorticity is not significant.  
 \item Gap opens in the vicinity of the planet.
\end{enumerate}
We investigate the shock formation process and the mass flux
separately.  Shock formation processes are investigated by Goodman and
Rafikov (2001) and Rafikov (2002a) using Burgers equation model, 
and the mass flux is investigated by Lubow (1990), using second-order
perturbation theory.  We show how these two theories can be combined to
investigate the gap opening processes. 

Physically, it is natural that the gap opens when spiral density wave
damps, since the angular momentum flux carried by the density wave
should be transferred to the background flow.  However, we shall show
that the decay of the spiral density wave away from the planet can
affect the mass flux in the immediate vicinity of the planet, in
contrast to the model suggested by Rafikov (2002b), in which there is no
gap formation in the vicinity of the planet in inviscid cases.

\subsection{Linear Analysis}
\label{subsec:linear}

In this section, as a preparation for the non-linear analyses performed
in the subsequent sections, 
we briefly summarize the results of linear study of the
spiral density wave (e.g., Goldreich and Tremaine 1979, 1980).  
We denote background state by subscript ``0'', and the perturbation by
$\delta$: 
\begin{equation}
 \Sigma = \Sigma_0 + \delta \Sigma,
\end{equation}
\begin{equation}
 \mathbf{v} = \mathbf{v}_0 + \delta \mathbf{v} = -\dfrac{3}{2}
  \Omega_{\rm p} x \mathbf{e}_y + \delta \mathbf{v} .
\end{equation}
The planet potential is regarded as a perturbation.  The linearized
equations are
\begin{equation}
 \left( \dfrac{\partial}{\partial t} - \dfrac{3}{2}\Omega_{\rm p}
  x \dfrac{\partial}{\partial y} \right) \dfrac{\delta\Sigma}{\Sigma_0}
 + \dfrac{\partial}{\partial x} \delta v_x
 + \dfrac{\partial}{\partial y} \delta v_y = 0
 \label{EoC_linpert}
\end{equation}
\begin{equation}
 \left( \dfrac{\partial}{\partial t} - \dfrac{3}{2}\Omega_{\rm p}
 x  \dfrac{\partial}{\partial y} \right) \delta v_x =
 -c^2 \dfrac{\partial}{\partial x} \dfrac{\delta \Sigma}{\Sigma_0}
 + 2\Omega_{\rm p} \delta v_y - \dfrac{\partial}{\partial x}
 \psi_{\mathrm p}
 \label{EoMx_linpert}
\end{equation}
\begin{equation}
 \left( \dfrac{\partial}{\partial t} - \dfrac{3}{2}\Omega_{\rm p}
 x  \dfrac{\partial}{\partial y} \right) \delta v_y =
 -c^2 \dfrac{\partial}{\partial y} \dfrac{\delta \Sigma}{\Sigma_0}
 - \dfrac{1}{2} \Omega_{\rm p} \delta v_x 
 - \dfrac{\partial}{\partial y} \psi_{\mathrm p}
 \label{EoMy_linpert}
\end{equation}
We now assume the stationary state in the frame corotating with the
planet, $\partial/\partial t=0$, and Fourier transform in the
$y$-direction.  The perturbed values are given by
\begin{equation}
 \delta f(x,y) = \sum_{n_y \in Z} \delta f(x) e^{i k_y y},
\end{equation}
where $\delta f$ denotes $\delta \Sigma$, $\delta v_x$, or $\delta v_y$,
and $k_y$ is the wave number in the $y$-direction. Assuming the
periodicity in the $y$-direction, $k_y=2\pi n_y/L_y$, where $n_y$ is an
integer and $L_y$ is the box size in the $y$-direction.  The summation
in the above equation is taken over the relative integers denoted by
$n_y$.   
It is possible to derive a single second-order ordinary differential
equation (Artymowicz 1993). 
\begin{equation}
 \dfrac{d^2}{dx^2} \delta v_y 
  + \left( \dfrac{9}{4} \dfrac{\Omega_{\rm p}^2 k_y^2}{c^2}x^2
    - k_y^2 - \dfrac{\Omega_{\rm p}^2}{c^2} \right) \delta v_y
  = \dfrac{3}{2} \dfrac{\Omega_{\rm p}k_y^2}{c^2} x \psi_{\rm p}
  - \dfrac{c^2 \Omega_{\rm p}}{2} \dfrac{d\psi_{\rm p}}{dx},
  \label{eq_dvy}
\end{equation} 
and other perturbed quantities are given by
\begin{equation}
 \dfrac{\delta\Sigma}{\Sigma_0} = \dfrac{1}{D} 
  \left[ \dfrac{\Omega_{\rm p}}{2} \dfrac{d}{dx} \delta v_y
 + \dfrac{3}{2} \Omega_{\rm p} k_y^2 x \delta v_y
 - k_y^2 \psi_{\rm p}
  \right]
  \label{eq_rho}
\end{equation}
and 
\begin{equation}
 \delta v_x = \dfrac{1}{D} 
  \left[ -c^2 ik_y \dfrac{d}{dx} \delta v_y
 + \dfrac{3}{4} \Omega_{\rm p}^2 ik_y x \delta v_y
 - i k_y \dfrac{\Omega_{\rm p}}{2} \psi_{\rm p}
  \right],
  \label{eq_dvx}
\end{equation}
where 
\begin{equation}
 D = \dfrac{\Omega_{\rm p}^2}{4} + c^2 k_y^2.
\end{equation}

The boundary condition for Equation \eqref{eq_dvy} is that wave should
propagate away from the planet.  At the location away from the effective
Lindblad resonance given by
\begin{equation}
 \dfrac{9}{4} \dfrac{\Omega_{\rm p}^2 k_y^2}{c^2}x^2
    - k_y^2 - \dfrac{\Omega_{\rm p}^2}{c^2} = 0,
\end{equation}
the approximate solution can be written analytically using WKB
approximation
%%%%%%%%%%%%%%%%%%%%%%%%%%%%
\footnote{The exact solution of equation \eqref{eq_dvy} is given by
parabolic cylinder function, see Artymowicz (1993) for detail.}  
%%%%%%%%%%%%%%%%%%%%%%%%%%%%
.
It takes the form, for $|x|\to\infty$, 
\begin{equation}
 \delta v_y \sim \dfrac{C(k_y)}{\sqrt{|x|}} \exp 
  \left[ \pm i \dfrac{3}{4}\dfrac{\Omega_{\rm p} k_y}{c} x^2 
  \right], 
  \label{dvy_sol}
\end{equation}
where upper sign is for $x>0$ and the lower sign is for $x<0$, and 
$C(k_y)$ denotes the amplitude (but not necessarily real) 
of the wave with mode $k_y$.    
We note that since we consider the linear perturbation 
excited by the planet's gravitational potential at the origin of our
coordinate system, the amplitude $C(k_y)$ is proportional to the planet
mass $M_{\rm p}$.
  From here on, using the symmetry of the shearing-sheet, 
we only consider $x>0$
without loss of generality.
Using equations \eqref{eq_rho}, \eqref{eq_dvx}, and \eqref{dvy_sol}, we
can derive the asymptotic form for $\delta \Sigma$ and $\delta v_x$ 
at $|x|\to\infty$,
\begin{equation}
 \dfrac{\delta \Sigma}{\Sigma_0} = 
  \dfrac{C(k_y) ik_y \Omega_{\rm p} \sqrt{x}}{D} 
  \left[  
   \dfrac{3}{4} \dfrac{\Omega_{\rm p}}{c} - i \dfrac{3}{2}k_y
  \right]
  \exp 
  \left[ i \dfrac{3}{4}\dfrac{\Omega_{\rm p} k_y}{c} x^2 
  \right]
  \label{rho_sol}
\end{equation}
\begin{equation}
 \delta v_x = c \dfrac{\delta \Sigma}{\Sigma_0},
  \label{vx_sol}
\end{equation}
where we have retained only the leading terms in $x$ in equations
\eqref{eq_rho} and \eqref{eq_dvx}.
Note that, in the real space, solutions are given by 
\begin{equation}
 \delta f = \sum_{k_y} F(k_y) x^{q} \exp
  \left[
   ik_y \left( y + \dfrac{3}{4}\dfrac{x^2}{H} \right)
  \right],
  \label{WKB_gen_sol}
\end{equation}
where $H$ is the scale height $H = c/\Omega_{\rm p}$, 
$q=1/2$ for $\delta \Sigma$ and $\delta v_x$, $q=-1/2$ for
$\delta v_y$, and $F(k_y)$ is a function of $k_y$. 
Therefore, along the line
\begin{equation}
 y = - \dfrac{3}{4} \dfrac{x^2}{H} + y_0,
  \label{constphase_lin}
\end{equation}
where $y_0$ is constant,
perturbed quantities take the same values except for the dependence 
$x^{q}$.  Therefore, we can write the form of the WKB solution 
in real space,
\begin{equation}
 \delta \Sigma (x,y) = M_{\rm p} x^{1/2} f(y+(3/4)x^2/H) ,
  \label{lin_form_dens}
\end{equation}
\begin{equation}
 \delta v_x (x,y) = M_{\rm p} x^{1/2} g(y+(3/4)x^2/H) ,
  \label{lin_form_vx}
\end{equation}
\begin{equation}
 \delta v_y (x,y) = M_{\rm p} x^{-1/2} h(y+(3/4)x^2/H) ,
  \label{lin_form_vy}
\end{equation}
where we write the dependence on the planet mass explicitly, and 
$f$, $g$, and $h$ are the functions that determines the form of
the perturbation.  We note that from equation \eqref{vx_sol}, $f$ and
$g$ are proportional
in the place where WKB approximation is valid.  

\begin{figure}
 \plotone{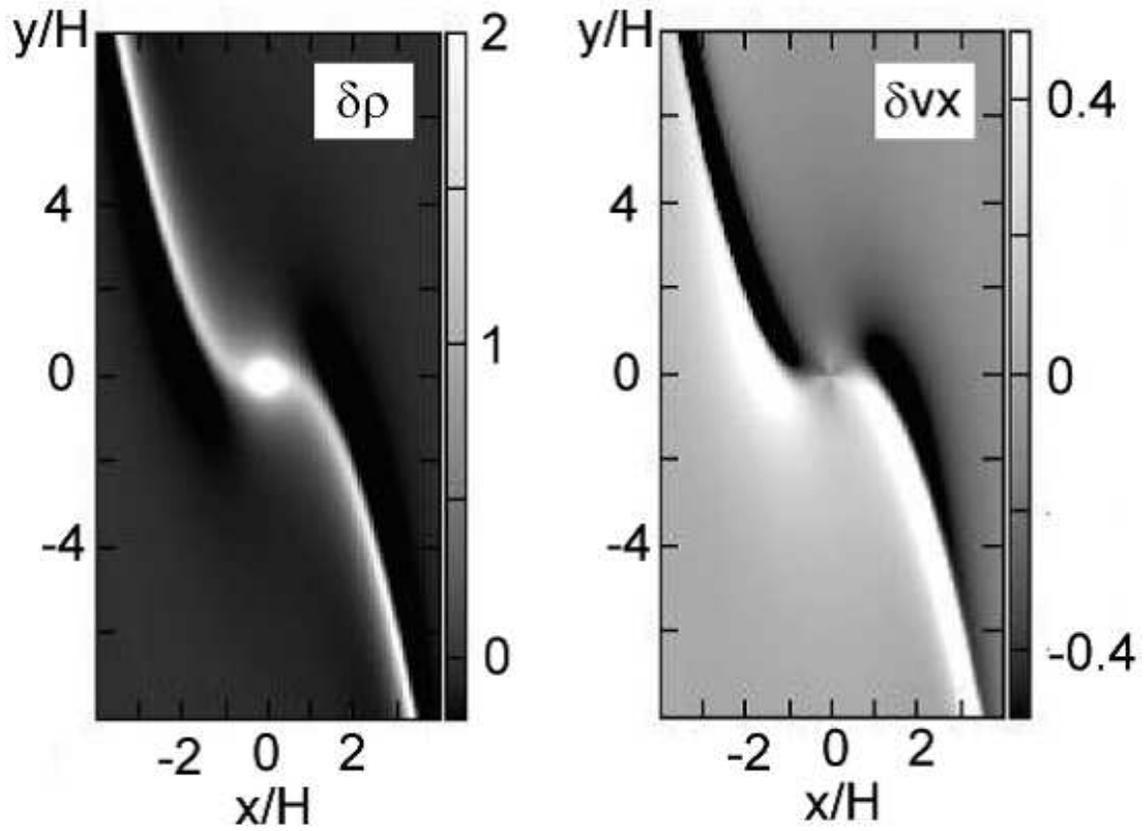}
 \caption{Linear density perturbation $\delta \Sigma/\Sigma_0$ (left)
 and $\delta v_x$ (right).  
 We use $GM_{\rm p}/Hc^2=1$, but the perturbation scales with the mass
 of the planet.
 }
 \label{fig:linear_profile}
\end{figure}

The exact linear solution can be obtained by solving equations
\eqref{EoC_linpert}-\eqref{EoMy_linpert} numerically with 
non-reflecting boundary conditions.  
The profiles of density and $v_x$ obtained in such a way are shown in
Figure \ref{fig:linear_profile}.
In this figure, we calculate non-axisymmetric modes 
($k_y\neq 0$) and assumed $GM_{\rm p}/Hc^2=1$.  The amplitude of
perturbation is proportional to the planet mass.   
We note that the results obtained in equations
\eqref{lin_form_dens}-\eqref{lin_form_vy} are based on the WKB
approximation, and they are applicable only in the region 
$|x/H| \gtrsim 1$.

\subsection{Shock Formation}
\label{subsec:shock}

Goodman and Rafikov (2001) performed a non-linear
analysis of the propagation of the spiral density wave, and their local
approach was later extended to the global model by Rafikov (2002a).    
Using several approximations based on the linear theory, 
they derived the
Burgers' equation which describes the propagation of spiral density
wave and concluded that the spiral density wave eventually
shocks as it propagates in the radial direction.  
They derived that the location of shock formation is
proportional to $M_{\rm p}^{-2/5}$.  In this section, we derive this
relationship using a slightly different consideration.

Shock is formed when two characteristics cross.  For two-dimensional
supersonic steady flow, the gradient of characteristic curves is given
by (Landau and Lifshitz 1959)
\begin{equation}
 \left( \dfrac{dy}{dx} \right)_{\pm} =
  \dfrac{v_x v_y \pm c \sqrt{v^2 - c^2}}{v_x^2 - c^2}.
\end{equation}
In the background state of the shearing-sheet, this gives
\begin{equation}
 \left( \dfrac{dy}{dx} \right)_{\pm} = 
  \mp \dfrac{1}{c}
  \left( \dfrac{9}{4} \Omega_{\rm p}^2 x^2 - c^2 \right)^{\frac{1}{2}} .
\end{equation}
For $x>0$, $(dy/dx)_{+}$ is the perturbation propagating away from the
origin.  At $x\gg (2/3)H$, the characteristic curves are given by
\begin{equation}
 y \sim - \dfrac{3}{4} \dfrac{x^2}{H} + y_0,
\end{equation}
where $y_0$ is a constant.  
It is to be noted that the outgoing characteristics coincides the curve
of the same phase of the perturbation of the density wave (see equation
\eqref{constphase_lin}).   This is because 
the density wave is essentially the sound wave
propagating on the disk.

For the flow distorted by the perturbation of the planet, the
characteristic curve is also distorted.  Using the results of 
linear perturbation, the gradient of the characteristics is given by 
\begin{equation}
 \left( \dfrac{dy}{dx} \right)_{\pm} \sim
  \dfrac{1}{c^2} 
  \left[
   \mp c \left( \dfrac{9}{4}\Omega_{\rm p}^2 x^2 - c^2
	 \right)^{\frac{1}{2}}
   + \dfrac{3}{2} \Omega_{\rm p} x 
   \left\{
    \delta v_x \pm 
    \dfrac{c\delta v_y}{\left( 
			 (9/4)\Omega_{\rm p}^2 x^2 - c^2 \right)^{1/2}}
   \right\}
  \right],
\end{equation}
upto the lowest order of perturbation. 
Since the amplitude of $\delta v_y$ decreases with $x^{-1/2}$, 
$\delta v_y$ in the second term of the right hand side can be
neglected 
for $|x/H|\gg 1$.  
Assuming that the perturbed  
characteristic curve is given by the form
\begin{equation}
 y(x) = - \dfrac{3}{4} \dfrac{x^2}{H} + y_0 + \delta y(x), 
\end{equation}
where $\delta y(x)$ is given by
\begin{equation}
 \dfrac{d}{dx} \delta y = \dfrac{3}{2}\dfrac{\Omega_{\rm p}}{c^2} x
  \delta v_x(x,y(x)) .
\end{equation}
Approximating $y(x) \sim -(3/2)x^2/H + y_0$ 
in the argument of $\delta v_x$, we have
\begin{equation}
 \delta y(x) \propto M_{\rm p} x^{5/2} g(y_0),
\end{equation}
where $g(y)$ is the function that appears in the linear solution of
$\delta v_x$ given by equation \eqref{lin_form_vx}.  From Figure
\ref{fig:linear_profile}, $g(y)$ has a zero point.  
In the vicinity of $g(y_0)=0$, $g(y)$ is negative for $y>y_0$ 
and $g(y)$ is positive for $y<y_0$.  This indicates that 
the separation between two 
characteristic curves shrinks compared to the unperturbed case.
Assuming that when $\delta y$ exceeds a critical value, the shock forms,
we obtain, for the location of the shock formation,
\begin{equation}
 x \propto M_{\rm p}^{-2/5},
\end{equation} 
which is the same condition as derived by Goodman and Rafikov (2001).  
Since the flow is supersonic only at $x>(2/3)H$, condition
\begin{equation}
 \left| x-\dfrac{2}{3}H \right| \propto M_{\rm p}^{-2/5}
  \label{shockform_cond}
\end{equation}
may be more appropriate.

\subsection{Second-Order Perturbation Analysis and Mass Flux}
\label{subsec:second}

In this section, we consider second-order linear perturbation in order
to derive the mass flux in the vicinity of the planet.  Although angular
momentum flux can be calculated using the results of linear perturbation
only, it is necessary to perform second-order analysis to derive the
mass flux, since axisymmetric ($k_y=0$) mode of the second-order
perturbation contributes to the mass flux.  Lubow (1990) performed the
time-dependent analysis using Fourier and Laplace transformation.  In
this paper, we calculate second order perturbation in real space.  

The equations for second-order perturbation are given by
\begin{eqnarray}
& \left( \dfrac{\partial}{\partial t} - \dfrac{3}{2}\Omega_{\rm p}
  \dfrac{\partial}{\partial y} \right) 
 \dfrac{\delta\Sigma^{(2)}}{\Sigma_0}
& + \dfrac{\partial}{\partial x} \delta v_x^{(2)}
 + \dfrac{\partial}{\partial y} \delta v_y^{(2)} = \nonumber \\ 
&&  - \dfrac{\partial}{\partial x} \left( 
	\dfrac{\delta \Sigma^{(1)}}{\Sigma_0} \delta v_x^{(1)}
				 \right)
  - \dfrac{\partial}{\partial y} \left( 
	\dfrac{\delta \Sigma^{(1)}}{\Sigma_0} \delta v_y^{(1)}
				 \right)
 \label{EoC_2ndpert}
\end{eqnarray}
\begin{eqnarray}
& \left( \dfrac{\partial}{\partial t} - \dfrac{3}{2}\Omega_{\rm p}
  \dfrac{\partial}{\partial y} \right) \delta v_x^{(2)} 
 &+ c^2 \dfrac{\partial}{\partial x} \dfrac{\delta \Sigma^{(2)}}{\Sigma_0} 
 - 2\Omega_{\rm p} \delta v_y^{(2)} =  \nonumber \\
&& - \delta v_x^{(1)} \dfrac{\partial}{\partial x} \delta v_x^{(1)}
 - \delta v_y^{(1)} \dfrac{\partial}{\partial y} \delta v_x^{(1)}
 + c^2 \dfrac{\delta \Sigma^{(1)}}{\Sigma_0} \dfrac{\partial}{\partial x} 
 \dfrac{\delta \Sigma^{(1)}}{\Sigma_0}
 \label{EoMx_2ndpert}
\end{eqnarray}
\begin{eqnarray}
& \left( \dfrac{\partial}{\partial t} - \dfrac{3}{2}\Omega_{\rm p}
  \dfrac{\partial}{\partial y} \right) \delta v_y^{(2)} 
& + c^2 \dfrac{\partial}{\partial y} \dfrac{\delta \Sigma^{(2)}}{\Sigma_0} 
 + \dfrac{1}{2}\Omega_{\rm p} \delta v_x^{(2)} = \nonumber \\
&& - \delta v_x^{(1)} \dfrac{\partial}{\partial x} \delta v_y^{(1)}
 - \delta v_y^{(1)} \dfrac{\partial}{\partial y} \delta v_y^{(1)}
 + c^2 \dfrac{\delta \Sigma^{(1)}}{\Sigma_0} \dfrac{\partial}{\partial x} 
 \dfrac{\delta \Sigma^{(1)}}{\Sigma_0}.
 \label{EoMy_2ndpert}
\end{eqnarray}
The superscripts ``$(1)$'' and ``$(2)$'' denote the first- and
second-order perturbation, respectively.  We assume that the first-order
results are already known.  

The mass flux is given by
\begin{equation}
 \mathcal{F}_M(t,x) = 
   \Sigma_0 \overline{\delta v_x^{(2)}} + 
   \overline{\delta \Sigma^{(1)} \delta v_x^{(1)}} ,
\end{equation}
where bar denotes the integral over $y$.  
Assuming the periodic boundary condition in the $y$-direction, it is
possible to derive the equation for $\mathcal{F}_M$ which reads
\begin{equation}
 \dfrac{\partial^2 \mathcal{F}_M}{\partial t^2} 
  - c^2 \dfrac{\partial^2 \mathcal{F}_M}{\partial x^2} 
  + \Omega_{\rm p}^2 \mathcal{F}_M = S(t,x), 
  \label{eq_mdot}
\end{equation}
where $S$ is the source term consisting of two parts,
\begin{equation}
 S(t,x) = S_v(t,x) + \dfrac{\partial}{\partial t} S_t(t,x),
\end{equation}
where
\begin{equation}
S_v(t,x) = 2 \Omega_{\rm p}
 \left[  
  \dfrac{\Omega_{\rm p}}{2} 
  \overline{\delta \Sigma^{(1)} \delta v_x^{(1)}}  
  -  \Sigma_0 
  \overline{\delta v_x^{(1)} \partial_x \delta v_y^{(1)}}   
 \right]
 \label{source_vort}
\end{equation} 
and
\begin{equation}
 S_t(t,x) = \dfrac{\partial}{\partial t} 
 \left( \overline{\delta \Sigma^{(1)}\delta v_x^{(1)}} \right)
 - \left[ \overline{\delta v_x^{(1)} \partial_x \delta v_x^{(1)}}
    + \overline{\delta v_y^{(1)} \partial_y \delta v_x^{(1)}}
    - \dfrac{c^2}{\Sigma_0} 
    \overline{ \delta\Sigma^{(1)} \partial_x \delta \Sigma^{(1)}} 
   \right]
 \label{source_time}
\end{equation} 
The term $S_v$ is related to the formation of specific vorticity.  In
two-dimensional ideal flow in a rotating frame, 
the specific vorticity conserves along the streamline 
\begin{equation}
 \dfrac{d}{dt} \dfrac{(\nabla \times \mathbf{v})_z 
  + 2\Omega_{\rm p}}{\Sigma} = 0,
\end{equation}
where 
$d/dt \equiv \partial_t + \mathbf{v}\cdot\nabla$ 
is the Lagrangian derivative.  
In the linear perturbation analyses, this reduces to
$\partial_t -(3/2)\Omega_{\rm p}\partial_y$, 
 see also equations \eqref{EoC_2ndpert}-\eqref{EoMy_2ndpert}.
The background value of the
specific vorticity is $\Omega_{\rm p}/2\Sigma_0$, 
and the linear perturbation is
\begin{equation}
 \left( \dfrac{\partial}{\partial t} - \dfrac{3}{2}\Omega_{\rm p}
  \dfrac{\partial}{\partial y} \right) 
  \left[
   \dfrac{\partial}{\partial y} \delta v_x^{(1)} - 
   \dfrac{\partial}{\partial x} \delta v_y^{(1)} + 
   \dfrac{1}{2} \Omega_{\rm p} \dfrac{\delta \Sigma^{(1)}}{\Sigma_0}
  \right] = 0 .
 \label{pert_vort}
\end{equation}
If there is no formation of specific vorticity, 
\begin{equation}
   \dfrac{\partial}{\partial y} \delta v_x^{(1)} - 
   \dfrac{\partial}{\partial x} \delta v_y^{(1)} + 
   \dfrac{1}{2} \Omega_{\rm p} \dfrac{\delta \Sigma^{(1)}}{\Sigma_0}
   = 0 .
\end{equation}
From equation \eqref{source_vort}, $S_v(t,x)$ becomes a total derivative
with respect to $y$ in this case, and therefore $S_v=0$.  
However, if there is a formation of vorticity by, for example, shock
damping of the spiral density wave, this term can not be neglected.  We
also note that if stationary state is assumed a priori, and if there is
no formation of specific vorticity, the source term $S(t,x)$ is zero,
leading to the zero mass flux (Lubow 1990, Muto and Inutsuka 2009a).

However, if time-evolution effects are taken into account, we have
non-zero mass flux.  The solution for equation \eqref{eq_mdot} is given
by
\begin{equation}
 \mathcal{F}_M(t,x) = \dfrac{1}{c} \int \int dt_0 dx_0 
  G(t,t_0;x,x_0) S(t_0,x_0),
  \label{mdot_sol}
\end{equation}
where $G(t,t_0;x,x_0)$ is the Green's function
\begin{eqnarray}
 \label{greenfunc}
 G(t,t_0;x,x_0) =
  \left\{
  \begin{array}{cc}
   \dfrac{1}{2} J_0\left( \dfrac{1}{H}\sqrt{c^2(t-t_0)^2 
		    - (x-x_0)^2} \right)   &
    |x-x_0|<c(t-t_0)
    \\[15pt] 
   0   &
    \mathrm{otherwise}
  \end{array}
  \right.,
\end{eqnarray}
where $J_0$ is the Bessel function of zeroth order. 
From linear perturbation, it is possible to predict that the mass flux
scales with $M_{\rm p}^2$, since the source term is the second order of
perturbation.  
Later in this paper, we compare this result with numerical calculations
to understand how much mass flux is excited by the planet.

We now consider the model in which the source term is given by
\begin{eqnarray}
 \label{source_model}
 S(t,x) =
  \left\{
  \begin{array}{cc}
   0 &    t<0
    \\[15pt] 
   S_0 \left[ \delta_D(x-x_s) - \delta_D(x+x_s) \right]  &
    t>0
  \end{array}
  \right.,
\end{eqnarray}
where $S_0>0$ and $x_s>0$ are positive constants, 
and $\delta_D(x)$ denotes the
Dirac's delta function.  We later see that the source term is positive
in the region $x>0$, and negative for $x<0$.  This form of the source
term is the simplest case where we can obtain an analytic solution for
the mass flux.

Changing the integration variable from $(t_0,x_0)$ to $(r,\theta)$ via
\begin{equation}
 x-x_0 = r \cos\theta
\end{equation}
\begin{equation}
 \dfrac{1}{H}\sqrt{ c^2(t-t_0)^2 - (x-x_0)^2 } = \dfrac{r}{H}
  \sin\theta, 
\end{equation}
equation \eqref{mdot_sol} can be rewritten to 
\begin{equation}
 \mathcal{F}_M(t,x) = \dfrac{H}{2c^2} \int_0^{ct} dr \int_0^{\pi} 
  d\theta
  \sin\theta J_0\left(\dfrac{r}{H}\sin\theta\right) 
  S\left[ t-\dfrac{r}{c}, x-r\cos\theta \right] .
\end{equation}
In order to see the mass flux only in the vicinity of the planet, we
assume $0<x<x_s$. 
Substituting equation \eqref{source_model}, and integrate over $r$, 
we obtain  
\begin{eqnarray}
& \mathcal{F}_M(t,x) = -\dfrac{HS_0}{2c^2} &
  \Bigg\{
   \int_{\cos^{-1}[(x-x_s)/ct]}^{\pi} d\theta 
   \left[
    \tan\theta J_0\left( \dfrac{x-x_s}{H}\tan\theta \right)
   \right]
   \nonumber  \\ 
&& +
      \int_{0}^{\cos^{-1}[(x+x_s)/ct]} d\theta 
   \left[
    \tan\theta J_0\left( \dfrac{x+x_s}{H}\tan\theta \right)
   \right]
  \Bigg\}
\end{eqnarray}
We take the limit $t\to\infty$.  Then, we can
approximate $\cos^{-1}[(x \pm x_s)/ct]\sim\pi/2$.  
Using the formula (Abramowitz and Stegun 1970)
\begin{equation}
 \int_{0}^{\pi/2} d\theta \tan\theta J_0 (a \tan\theta) 
  = \int_{0}^{\infty} du \dfrac{u}{1+u^2} J_0(au) = K_0 (a),
\end{equation}
where $a>0$, the integration
over $\theta$ can be performed to obtain
\begin{equation}
 \mathcal{F}_M(t,x) \sim \dfrac{HS_0}{2c^2} 
  \left[ 
   K_0\left( \dfrac{x_s-x}{H} \right) 
   - K_0 \left( \dfrac{x_s+x}{H} \right)
  \right],
  \label{mdot_model_sol}
\end{equation}
where $K_0$ is the modified Bessel function of the zeroth order.  
In the vicinity of the planet, $x \ll x_s$, the mass
flux changes with $\mathcal{F}_M\propto x$, which indicates that the gap opens
up.  We note that divergence at $x= \pm x_s$ is the artefact of our
simplification where we have used delta function as a source term.

\subsection{Instability of a Disk with Surface Density Variation}
\label{subsec:gapinst}

We have seen that if there is a source of specific vorticity, mass flux
appears in the vicinity of the planet, and it leads to the change of
surface density to open a gap.  
We now briefly discuss how this gap opening process ends.  

An inviscid disk
with a rapid surface density variation is prone to a linear instability, 
which is referred to as Rossby wave instability 
(Li et al. 2000, de Val-Borro et al. 2007). 
In Appendix \ref{app:RWI}, we show a brief outline of the linear
stability analyses for a disk with a gap, and derive the necessary
conditions for the instability.  
Gap induced by a planet in the disk naturally excites the variation of
specific vorticity, and gap edges are likely places where
such instability occurs.  Rossby wave instability 
may stop the gap-opening processes described in previous subsections.
We do not expect that the instability leads to the complete closing of
the gap, since the planet always try to repel the fluid element by the
excitation of the density wave.  There may be at least ``underdense
region'' around the planetary orbit, which may be called a ``gap''.

\subsection{Summary of Analytic Study}
\label{subsec:summary}

We have discussed the non-linear shock formation of density
wave in Section \ref{subsec:shock}.  The formation of shock leads to the
formation of specific vorticity in general, and the perturbation of
specific vorticity can lead to the radial mass flux as discussed in
Section \ref{subsec:second}.  The gap formation may end by the
onset of the linear instability of the disk with a gap.  
In order to investigate this qualitative picture of gap formation, we
perform two-dimensional numerical calculations in the
subsequent section.

We note that 
in this picture of gap formation, the decay of angular momentum flux
carried by the spiral density wave is not directly connected to the
mass flux.  For example, when we model the source of mass flux by
$\delta$-function in equation \eqref{source_model}, we implicitly assume
that the spiral density wave does not dissipate at $|x|<x_s$, but we
still end up with non-zero mass flux in this region, see equation
\eqref{mdot_model_sol}.   
One may consider that this result violates the conservation of
angular momentum.  However, it is actually satisfied, 
when time evolution effects are taken into
account.     
We shall discuss further on the angular momentum conservation in Section
\ref{subsec:1d}.

\section{Numerical Study of Planetary Wake}
\label{sec:numerical}

\subsection{Numerical Setup}
\label{subsec:method}

We solve Euler equations \eqref{EoC_full} and \eqref{EoM_full} with
isothermal equation of state using second-order Godunov scheme (e.g.,
Colella and Woodward 1984).  In order to investigate the density wave
propagation while resolving all the length scales (Hill radius, Bondi
radius, and radial wavelength of the density wave), 
we need to use rather large box size with relatively high resolution.
We choose the box size $(L_x,L_y)=(16H,32H)$ with mesh number 
$(N_x,N_y)=(512,1024)$, which results in the resolution 
$\Delta x=\Delta y=H/32$.
We have also used the box with different $L_y$ to investigate the box
size effect.  

From equation \eqref{WKB_gen_sol}, the radial wavelength of the wave
with mode $k_y$ may be approximated by
\begin{equation}
 \lambda \sim H \dfrac{8\pi}{3} \dfrac{1}{k_y H} \dfrac{H}{x} .
\end{equation}
The most important modes are $k_y H \sim \mathcal{O}(1)$.  For mode with
$k_y H=10$ at $x/H \sim 4$, the wavelength of the density wave is given
by $\lambda \sim H/4$, and therefore, the radial wavelength is
resolved by eight meshes  
if we use $\Delta x=H/32$.  Higher-order modes or the wave outside this
regions are damped numerically. 
Therefore, we mainly use data within $|x|<4H$ in the analyses of the
numerical results.     

If we normalize the length by $H=c/\Omega_{\rm p}$ and the
time by $\Omega_{\rm p}^{-1}$, 
the only dimensionless parameter in this
calculation is the planet mass, or Bondi radius 
$r_{\rm B}/H = GM_{\rm p}/Hc^2$ and the softening
parameter $\epsilon/H$.  We use the softening length used in previous
local shearing-sheet calculations $\epsilon=r_{\rm H}/4$, 
where $r_{\rm_H}$ is the Hill's radius
\begin{equation}
 \dfrac{r_{\rm H}}{H} 
  = \left( \dfrac{M_{\rm p}}{3M_{\rm \ast}} \right)^{1/3} 
     \dfrac{r_{\rm p}}{H} 
  = \left(\dfrac{GM_{\rm p}}{3Hc^2}\right)^{1/3}
\end{equation}
(Miyoshi et al. 1999).  
We have performed calculations 
with five different planetary mass $M_{\rm p}$
shown in Table \ref{table:parameter}.  Note that Bondi radius, Hill
radius, and the softening parameter are resolved (at least marginally)
for the smallest mass model.  We have increased the planet mass 
linearly from $t\Omega_{\rm p}=0$ to $12$.  The result does not depend
significantly on this timescale if the timescale is longer than
this. 

In the $x$-direction, 
we use non-reflecting boundary  
used by FARGO (Baruteau 2008), modified for the shearing-sheet.  
For the $y$-direction, we adopt the periodic boundary condition.
Shearing-sheet calculations performed by Miyoshi et al. (1999) 
or Tanigawa and Watanabe (2002) 
used a different boundary condition in the
$y$-direction, which is the combination of Keplerian inflow and
supersonic outflow.  
This boundary condition may be appropriate to study the flow structure
only in the vicinity of the planet, 
but for the study of gap formation, this boundary
condition is inappropriate because this forces the unperturbed gas 
flowing into the computational domain.  We also note that periodic
boundary condition in the $y$-direction is useful for the purpose of
comparison with the linear analyses, which assume the periodicity in the
$y$-direction.  
We have checked that our code reproduces the results of Miyoshi et
al. (1999) well when the same parameter and the boundary conditions are
used.  

\subsection{Results}
\label{subsec:result}

\subsubsection{Disk Structure and Evolution}

\begin{figure}
 \plotone{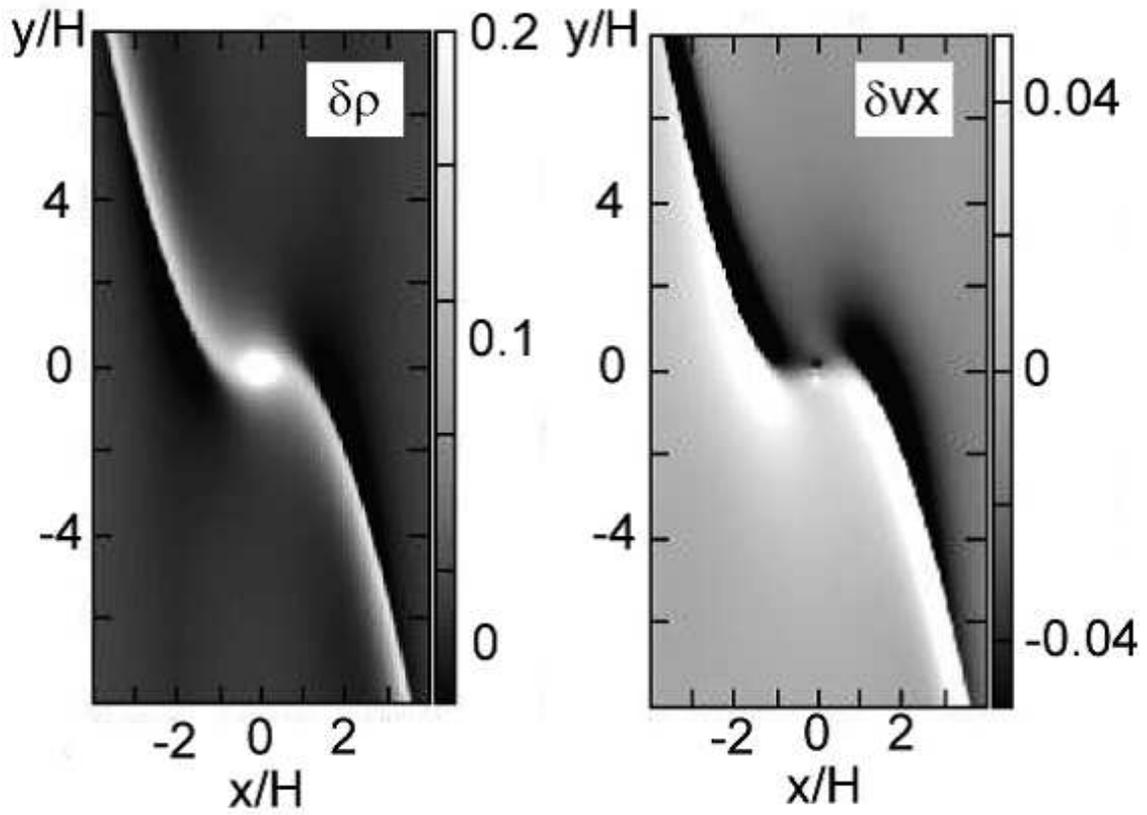}
 \caption{Simulation results of  $\delta \Sigma/\Sigma_0$ (left)
 and $\delta v_x$ (right) at $t\Omega_{\rm p}=200$.  
 We use $GM_{\rm p}/Hc^2=0.1$.
 }
 \label{fig:sim_M01_profile}
\end{figure}

\begin{figure}
 \plotone{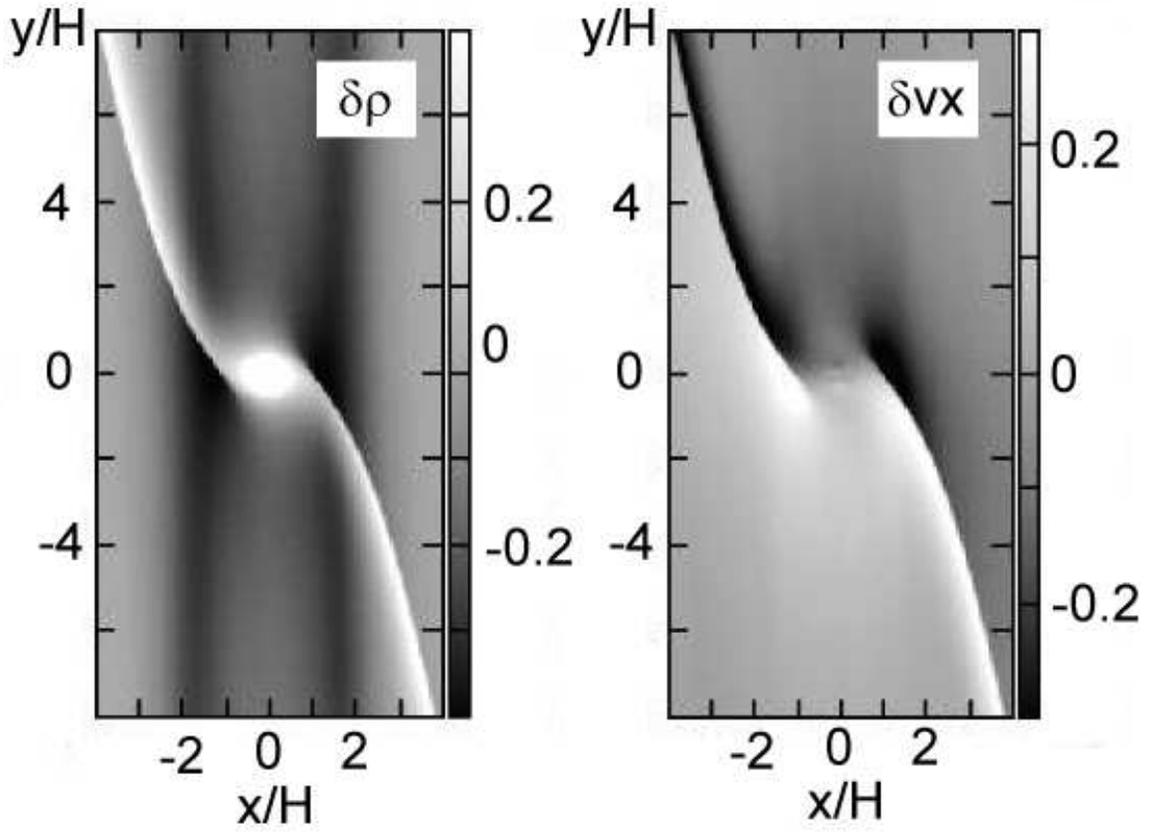}
 \caption{Simulation results of  $\delta \Sigma/\Sigma_0$ 
 and $\delta v_x$ at $t\Omega_{\rm p}=200$.  
 We use $GM_{\rm p}/Hc^2=0.4$.
 }
 \label{fig:sim_M04_profile}
\end{figure}

\begin{figure}
 \plottwo{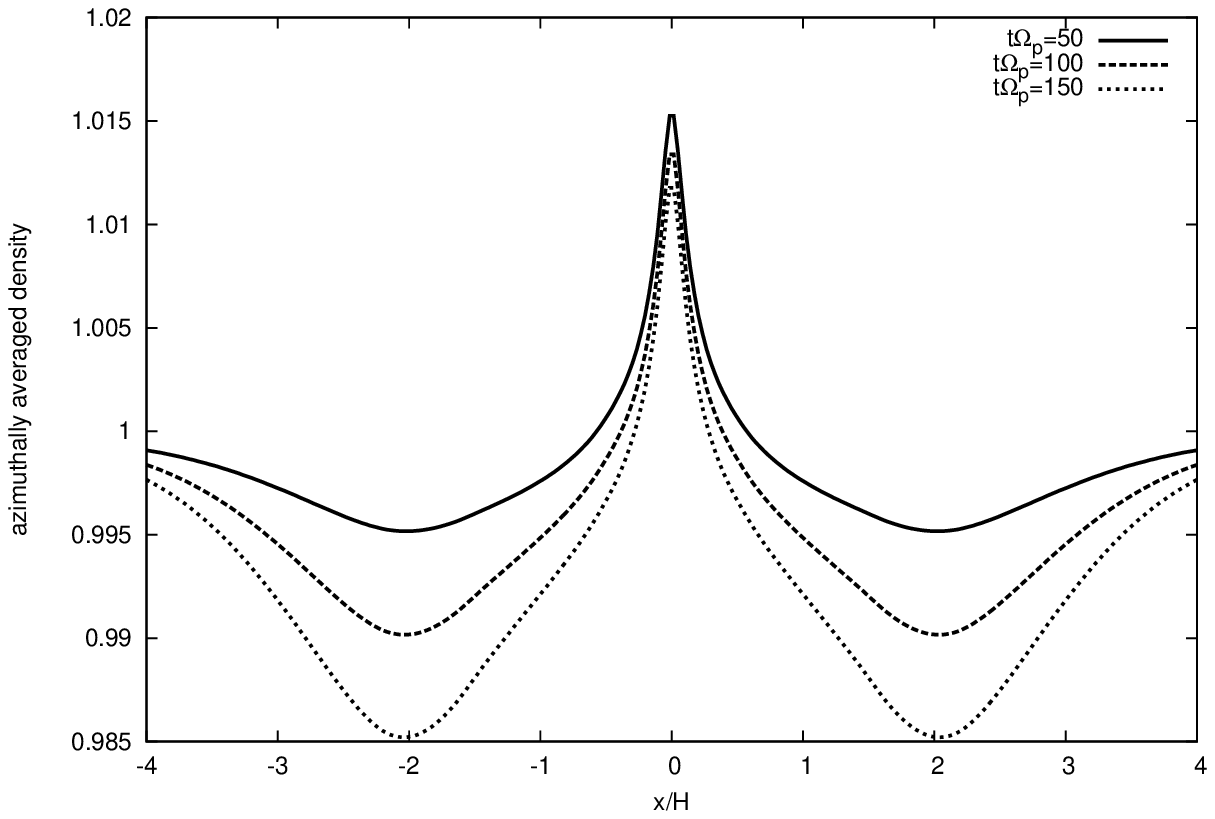}{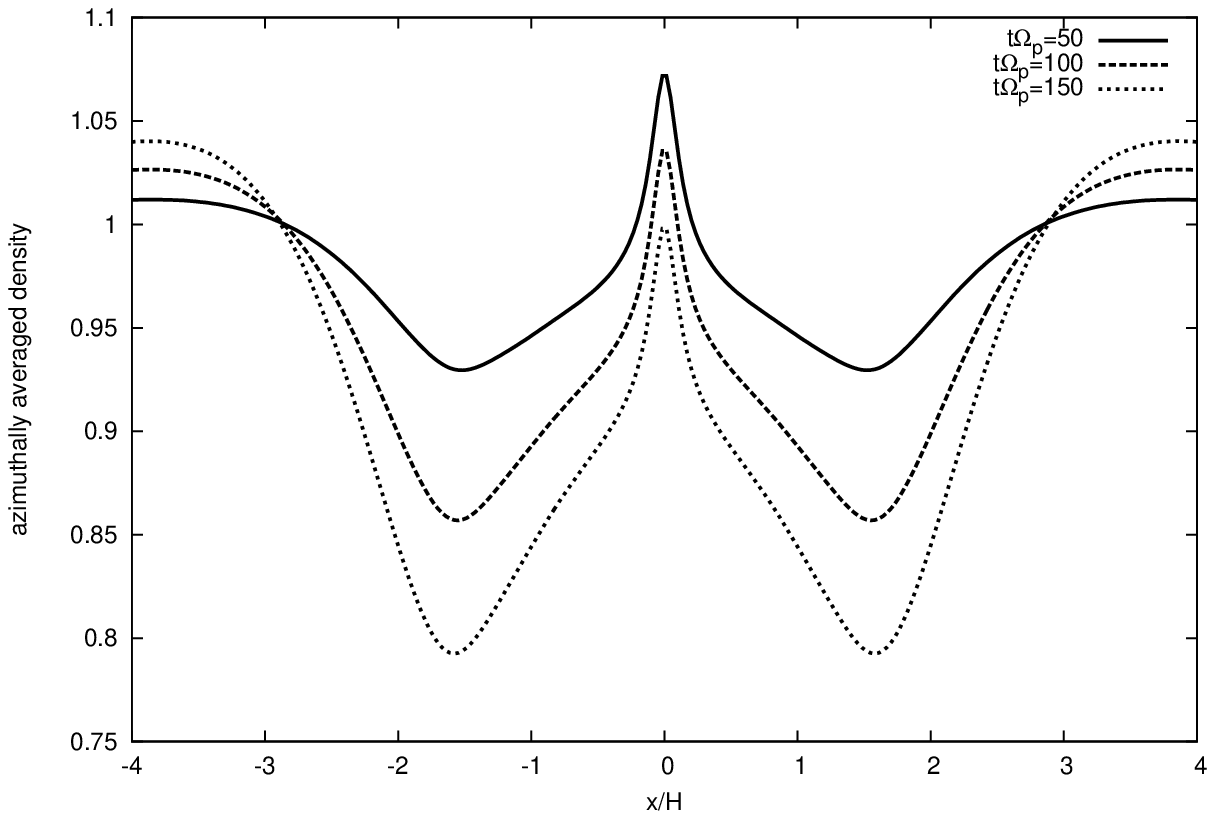}
 \caption{The evolution of azimuthally averaged density profile for
 $GM_{\rm p}/Hc^2=0.1$ (left) and $0.4$ (right).
 }
 \label{fig:gap_evolution}
\end{figure}

Figures \ref{fig:sim_M01_profile} and \ref{fig:sim_M04_profile} 
show the $\delta \Sigma$ and $\delta v_x$ at $t\Omega_{\rm p}=200$
 obtained by numerical simulations.  
For the numerical
calculations with $GM_{\rm p}/Hc^2=0.4$, we can see that a gap is
already formed.  
Figure \ref{fig:gap_evolution} shows the evolution of 
the azimuthally-averaged density profile for
$GM_{\rm p}/Hc^2=0.1$ and $0.4$.  It is possible to see that even for
low-mass calculations, density gap is being gradually formed.  
In subsequent 
sections, we provide the physical interpretations 
of disk structure and evolution using analytical theories provided in
Section \ref{sec:analytic}.

\subsubsection{Spiral Shock Wave}

\begin{figure}
 \plottwo{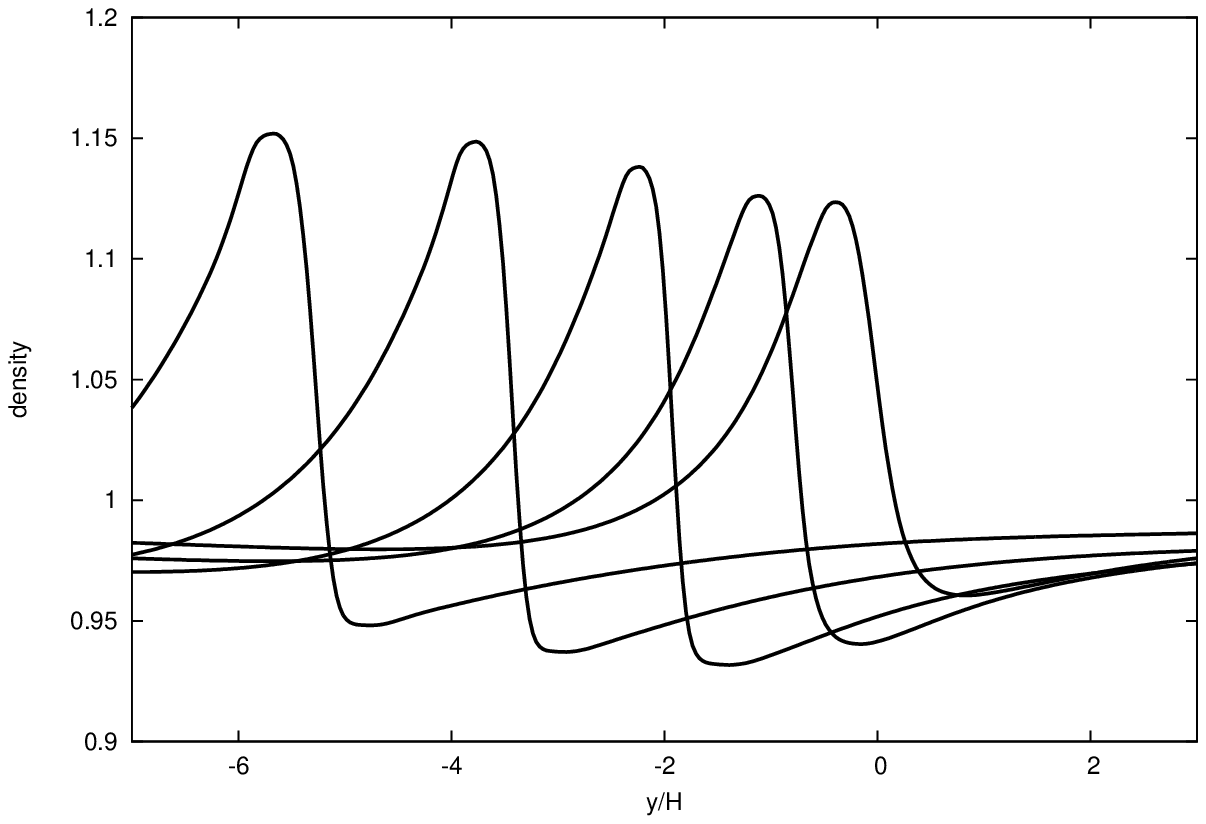}{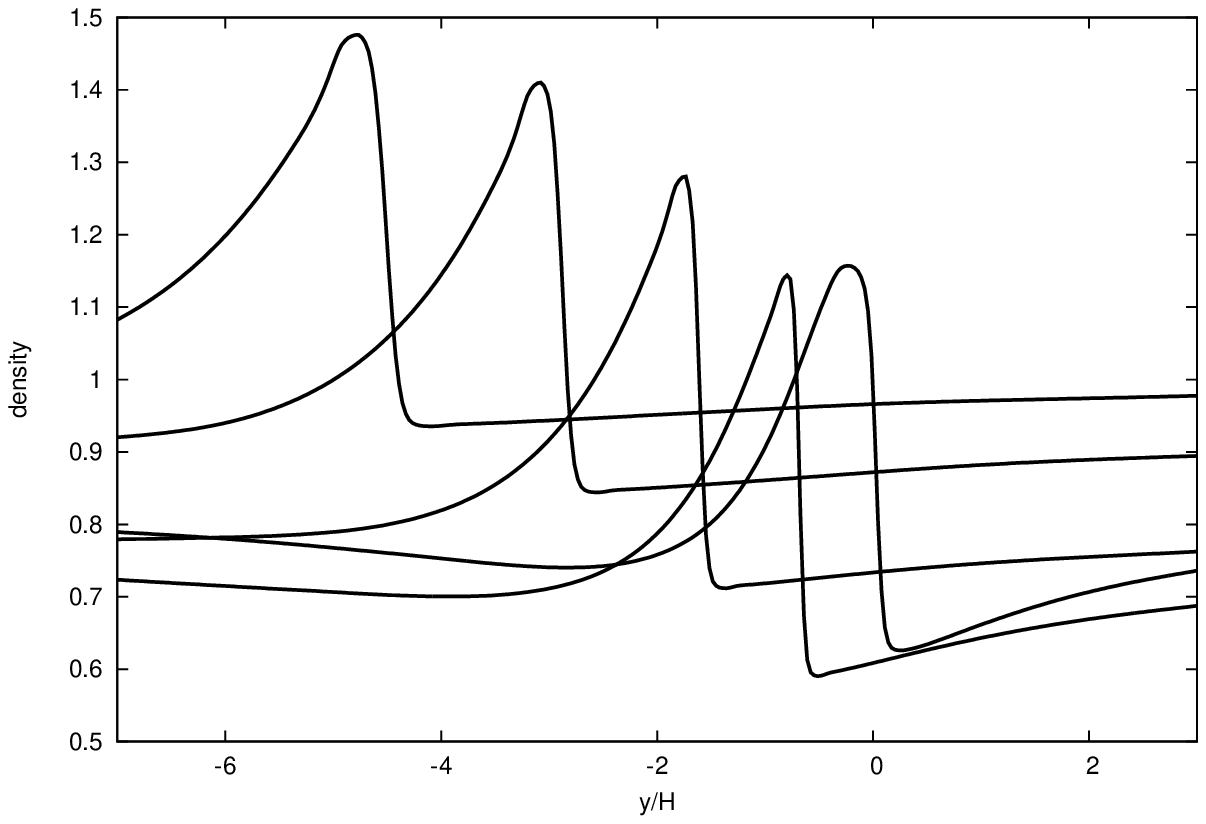}
 \caption{Density profile at $t\Omega_{\rm p}=200$.  Profiles at
 $x/H=1$, $1.5$, $2$, $2.5$, and $3$ are shown from right to left.  Left
 panel is for $GM_{\rm p}/Hc^2=0.1$ and the right panel is for 
 $GM_{\rm p}/Hc^2=0.4$.
 }
 \label{fig:densprof}
\end{figure}

We first investigate the shock formation.  
Figure \ref{fig:densprof} shows the density profiles at various $x$ at
$t\Omega_{\rm p}=200$ obtained by numerical calculations.
  It is possible to see the shock-like structure 
for the calculation with $GM_{\rm p}/Hc^2=0.4$, while for the run with 
$GM_{\rm p}/Hc^2=0.1$, the structure is not very obvious.

It is possible to see whether dissipation acts or not by looking at the
perturbation of specific vorticity.  
In the absence of dissipation, the specific vorticity conserves
along the streamline, and since the background 
specific vorticity is constant in the calculation box, we expect that
it is also constant in the presence of the planet.  Note that any
potential force, whether it is time-dependent or not, does not produce
specific vorticity.  Therefore, specific vorticity arises only if
dissipative mechanisms come into play.  
In our numerical calculation, the
dissipation is implemented in the shock-capturing scheme.  If the
formation of specific vorticity is driven by the formation of shock,
relation given by equation \eqref{shockform_cond} should be observed at
least for the weak shock cases.  

\begin{figure}
 \plottwo{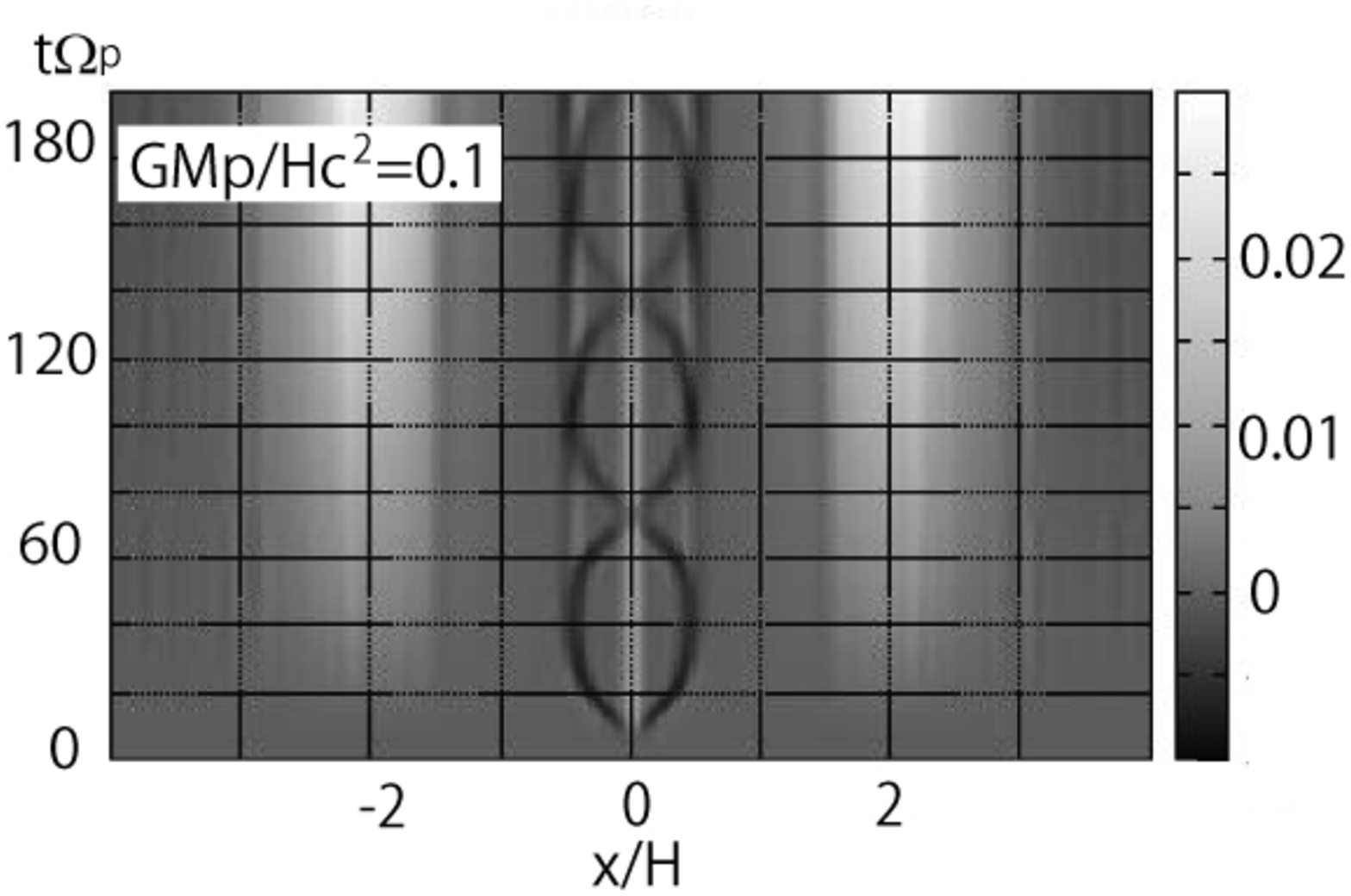}{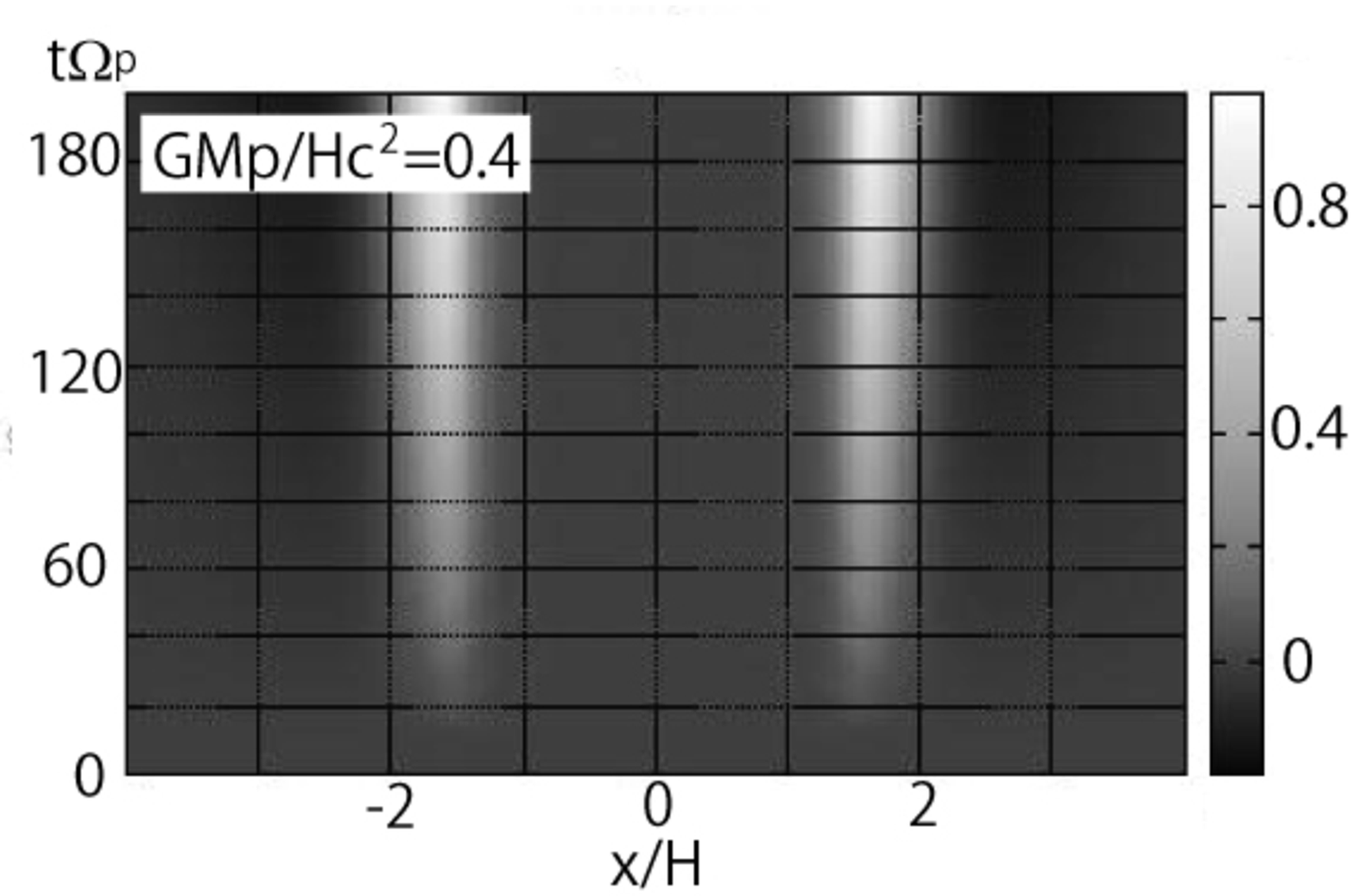}
 \caption{Time evolution of azimuthally averaged specific vorticity
 perturbation for $GM_{\rm p}/Hc^2=0.1$ (left) and $0.4$ (right). }
 \label{fig:vort_evol}
\end{figure}

Figure \ref{fig:vort_evol} shows the evolution of the perturbation of
azimuthally averaged specific vorticity for calculations with 
$GM_{\rm p}/Hc^2=0.1$ and $0.4$.  It is possible to see that the
specific vorticity is formed around $x/H\sim 1.5$.  
We later see that this specific vorticity becomes a dominant source of
mass flux, which leads to the formation of a gap.  

For the numerical calculation
with $GM_{\rm p}/Hc^2=0.1$, we also see the formation of specific
vorticity around $x \sim 0$.  This is because the fluid elements in the
vicinity of the planet orbit 
around it, and as a result, a vortex
is formed just around the planet.  The formation of a vortex at the
planet location causes the vortex circulating in the opposite direction,
which makes a horseshoe orbit in the corotation region.  For the
calculation with $GM_{\rm p}/Hc^2=0.4$, similar thing happens, but the
vortices formed by shock dissipation are stronger.  In subsequent
sections, we show that
vortices just in the vicinity of the planet location does not
produce a significant amount of mass flux.   

\begin{figure}
 \plotone{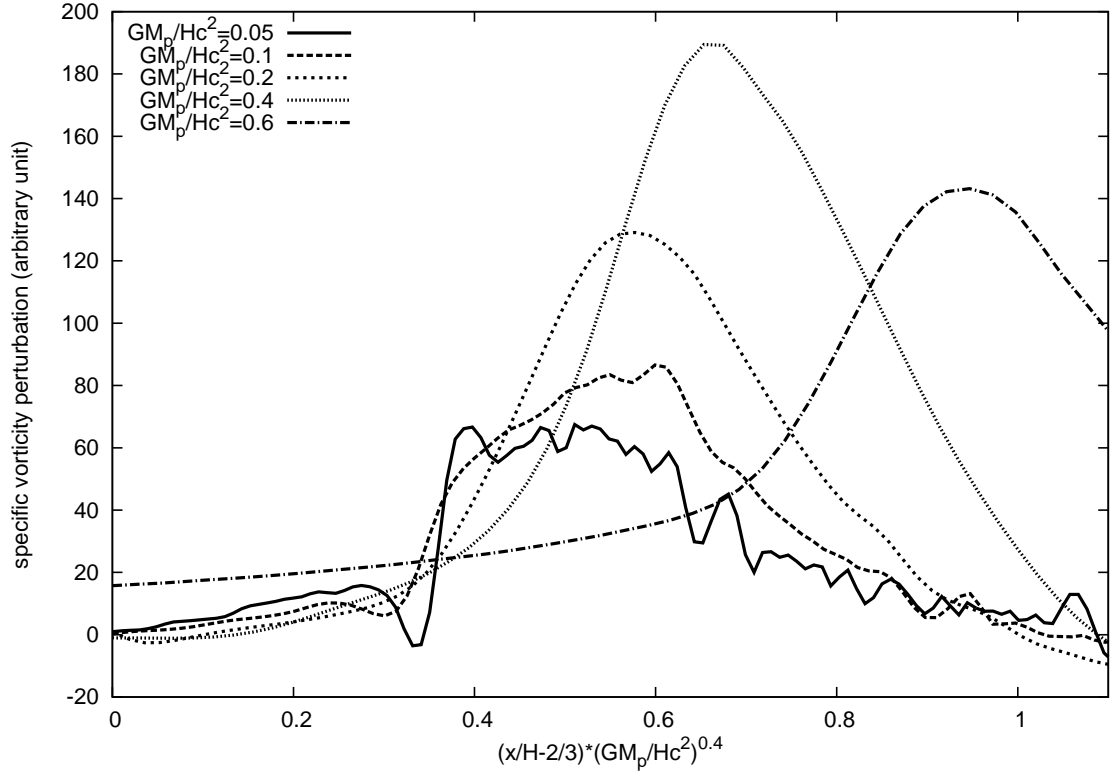}
 \caption{Perturbation of specific vorticity (in arbitrary unit) as a
 function of $(x-(2H/3)) \times M_{\rm p}^{2/5}$.
 }
 \label{fig:vort_pert}
\end{figure}

Figure \ref{fig:vort_pert} shows the azimuthally-integrated specific
vorticity perturbation as a function of 
$(x-(2/3)H)M_{\rm p}^{2/5}$.  
If shock dissipation occurs as equation \eqref{shockform_cond}, the peak
of the specific vorticity perturbation comes at the same location of the
horizontal axis.  It is possible to observe that the for calculations
with $GM_{\rm p}/Hc^2<0.2$, equation \eqref{shockform_cond} is
marginally satisfied.  For calculations with larger planet mass, it is
not the case.  This is because the shock formation occurs 
immediately after the excitation of the
 wave at $x\sim(2/3)H$, and therefore in
the unit of $(x-(2/3)H)M_{\rm p}^{2/5}$, 
the shock occurs relatively further away.

\subsubsection{Mass Flux}

\begin{figure}
 \plottwo{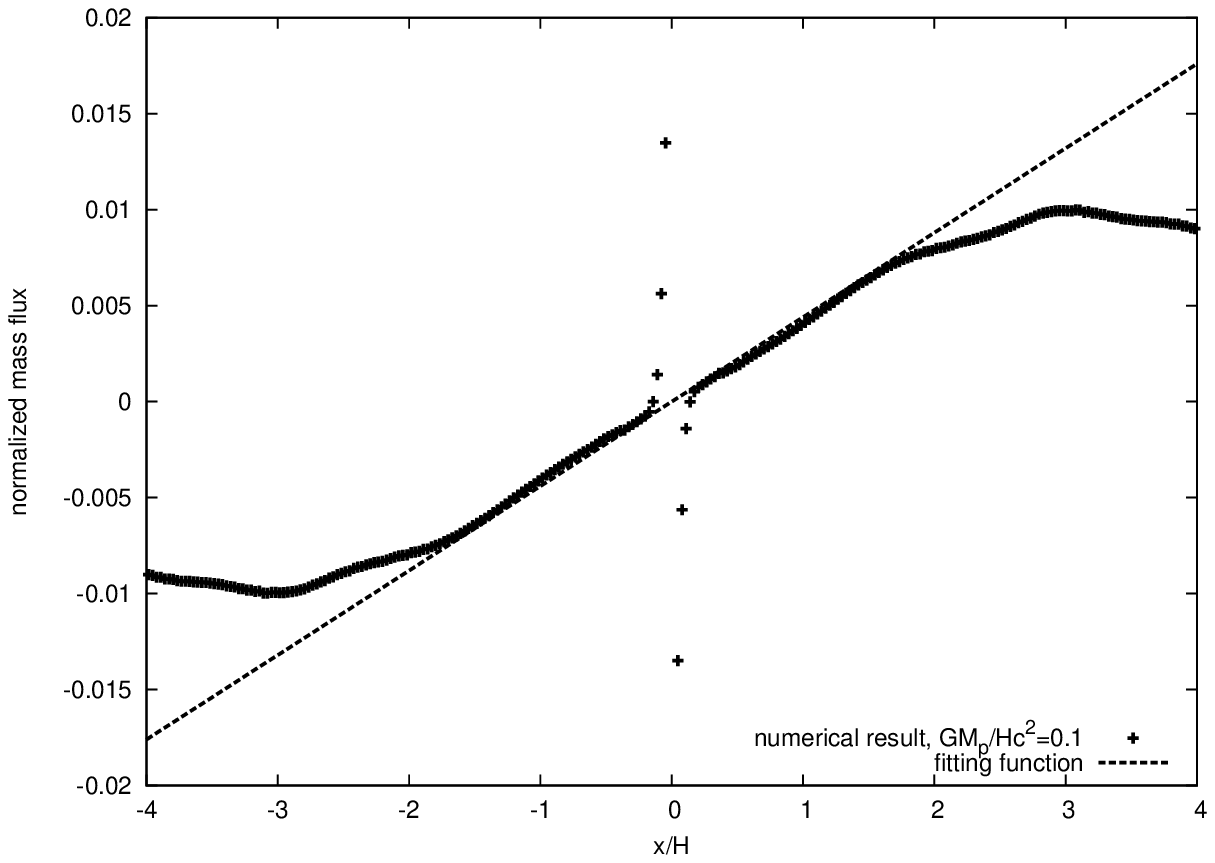}{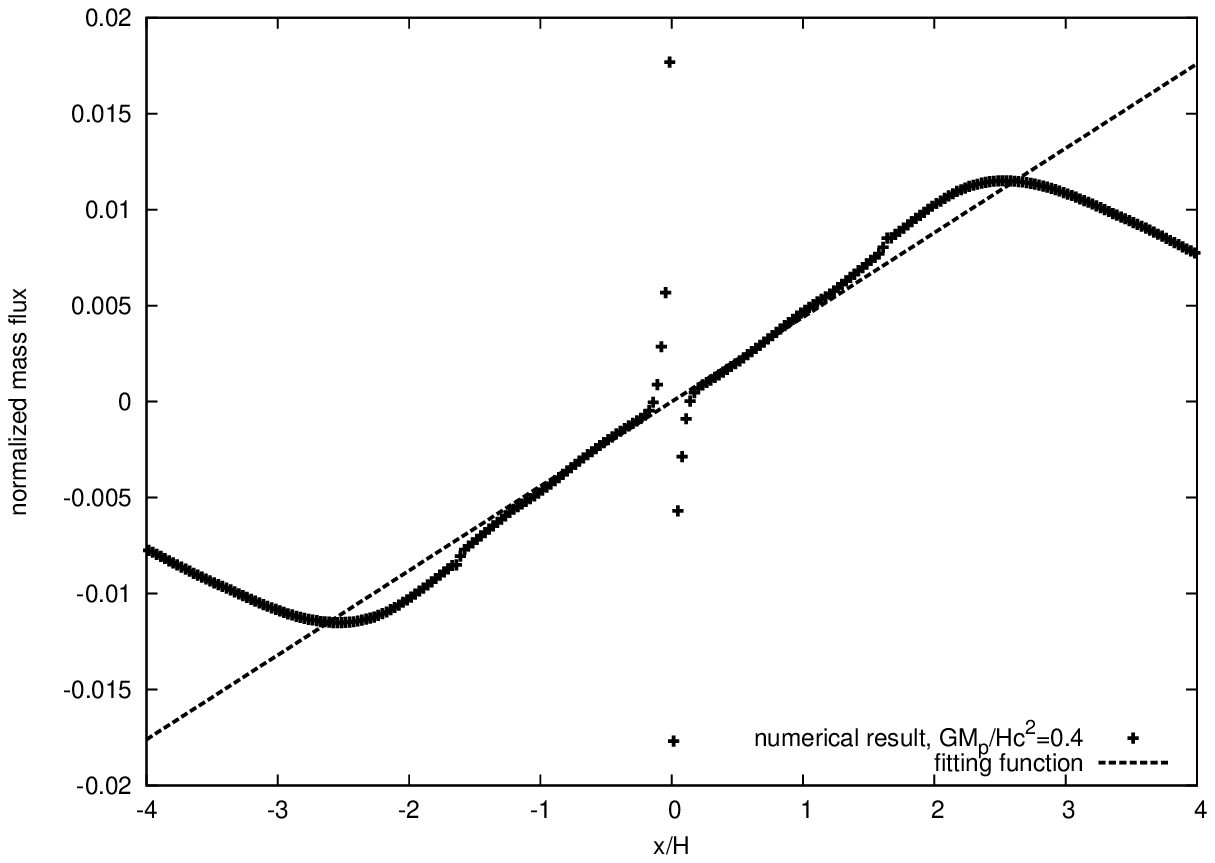}
 \caption{
 Comparison between the mass flux obtained from numerical calculation
 (dots) and the fitting function \eqref{massflux_fit} with
 $K=4.4\times10^{-3}$.  We plot the value 
 $(GM_{\rm p}/Hc^2)^{-2}(1/L_y)\int dy \Sigma v_x$.  Left panel shows the
 results with $GM_{\rm p}/Hc^2=0.1$ and the right panel shows the
 results with $GM_{\rm p}/Hc^2=0.4$.  The horizontal axis shows $x/H$.
 }
 \label{fig:mdot_fit}
\end{figure}

We now look at the radial mass flux $\overline{\Sigma v_x}$ excited by the
planet.  
In figure \ref{fig:mdot_fit}, we show the azimuthally averaged mass flux
obtained by numerical calculation.  It is possible to see that 
there is a spike of the mass flux just in the vicinity of the planet, 
and then the mass flux depends linearly with the distance from the
planet within $|x/H|\lesssim 2$.  The spike in the vicinity of the
planet is due to the gas falling onto the planet.
This causes the spike of the surface density profile in the vicinity
of the planet as shown in figure \ref{fig:gap_evolution}.  
We expect that this feature depends on the treatment of the planet, and
such mass flux should be investigated carefully when one wants to look
at the processes such as gas accretion onto the
planet.  However, as we will see later, the feature in this spike region
does not affect the mass flux outside this region.  In the
discussion below, we focus on the structure outside this spike region.

\begin{figure}
 \plottwo{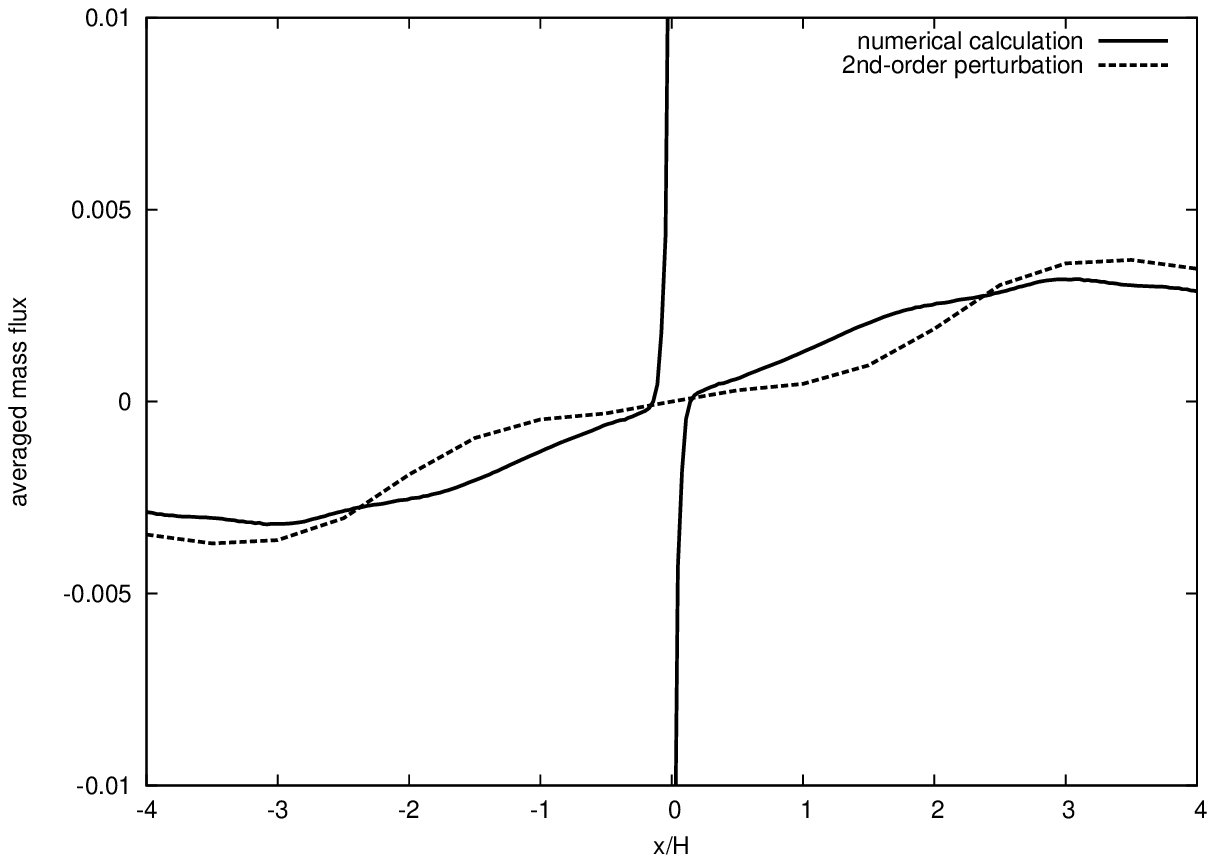}{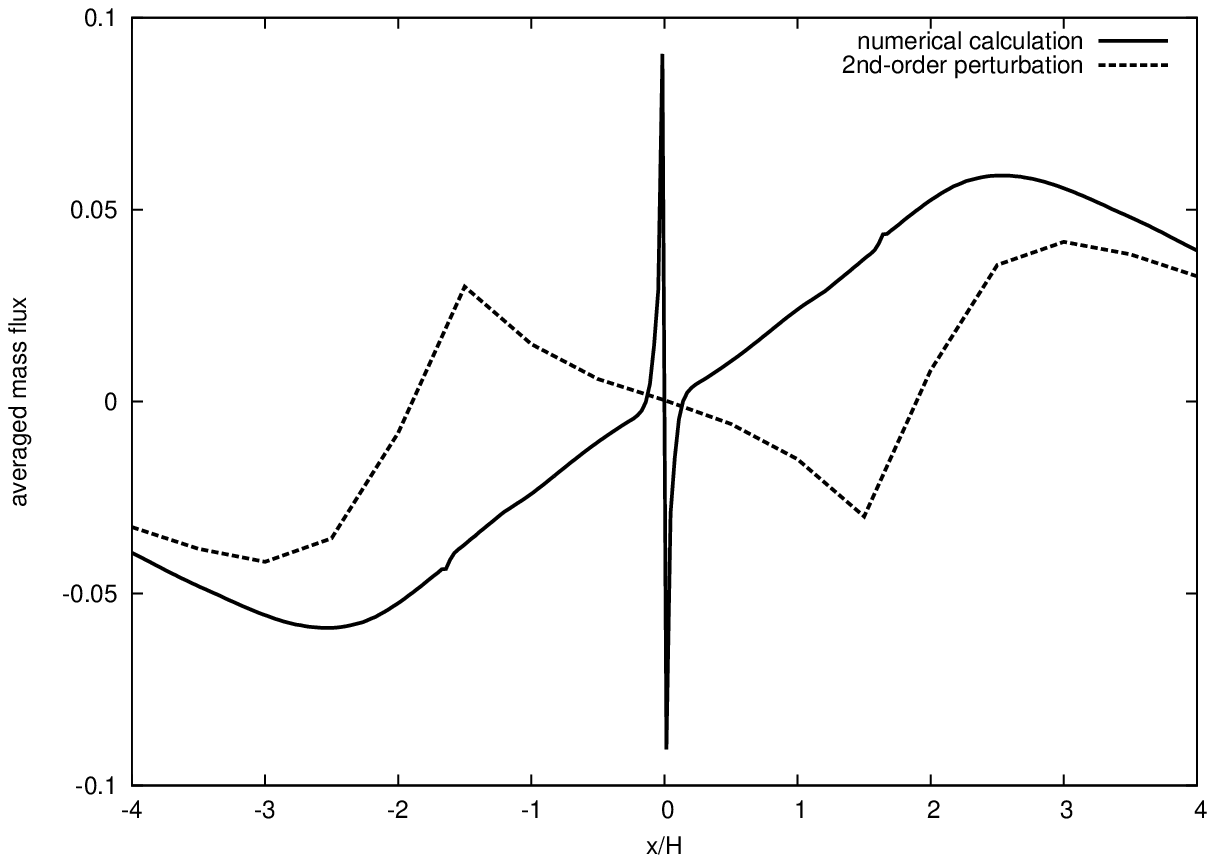}
 \caption{
 Comparison of the mass flux obtained by numerical calculation 
 and second-order perturbation theory.  Left panel shows the result with
 $GM_{\rm p}/Hc^2=0.1$, and the right panel shows the results with 
 $GM_{\rm p}/Hc^2=0.4$.
 }
 \label{fig:mdot_comp}
\end{figure}

We first compare the results obtained by numerical simulation and
analytic calculation.  
Figure \ref{fig:mdot_comp} compares the mass flux derived by numerical
calculation and by equation \eqref{mdot_sol}.  
Since it is not possible to predict the amount of specific vorticity
only from linear perturbation theory, we have used the results of
numerical calculation in obtaining  the source term $S(t,x)$.
It is possible to observe that the second-order perturbation theory is
valid for calculations with low-mass planet, while for calculations with
$GM_{\rm p}/Hc^2=0.4$, the perturbation theory fails to explain the
amount of mass flux.  We find that the second-order perturbation theory
can be used for $GM_{\rm p}/Hc^2 \lesssim 0.2$.  

\begin{figure}
 \plottwo{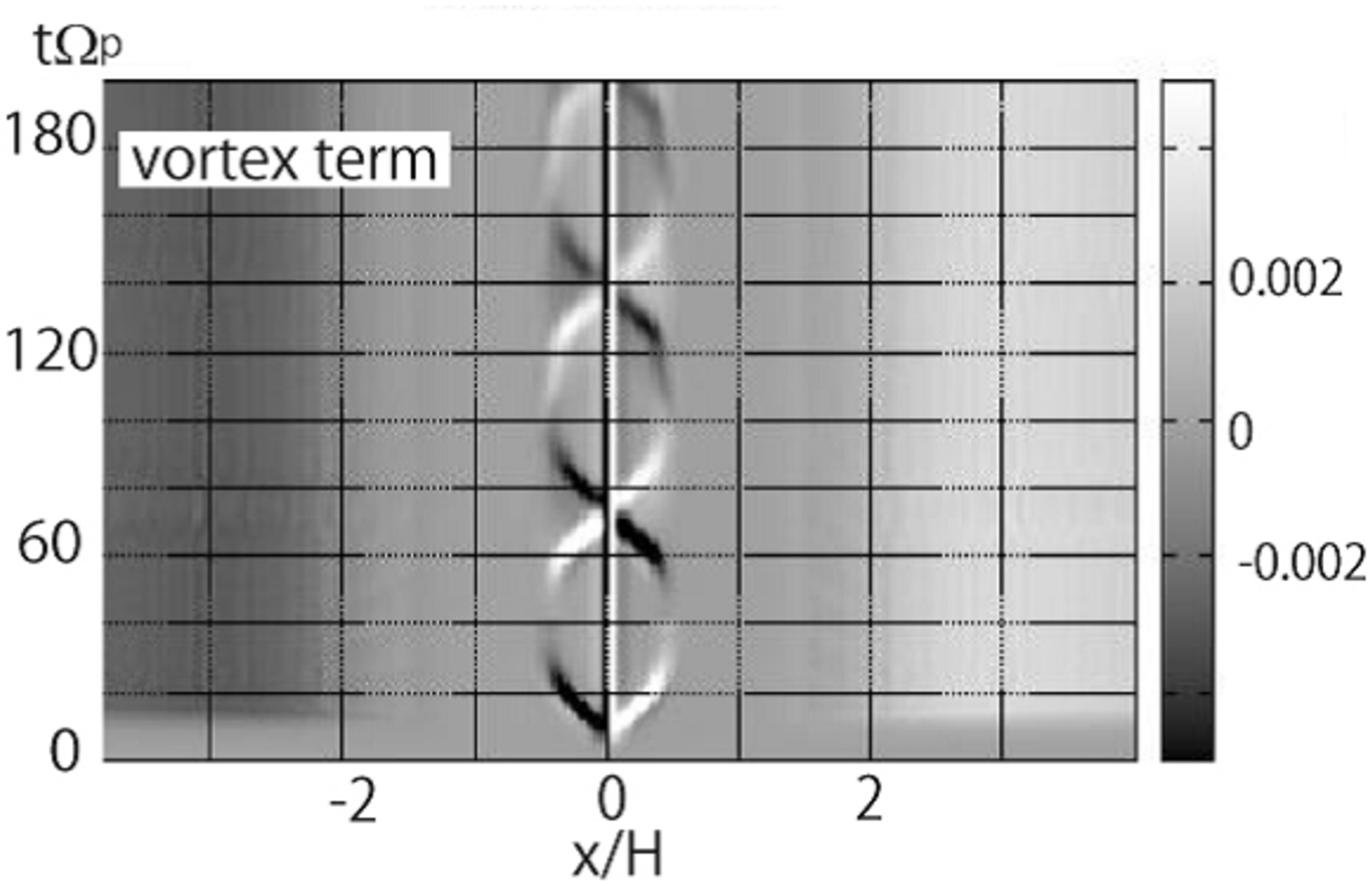}{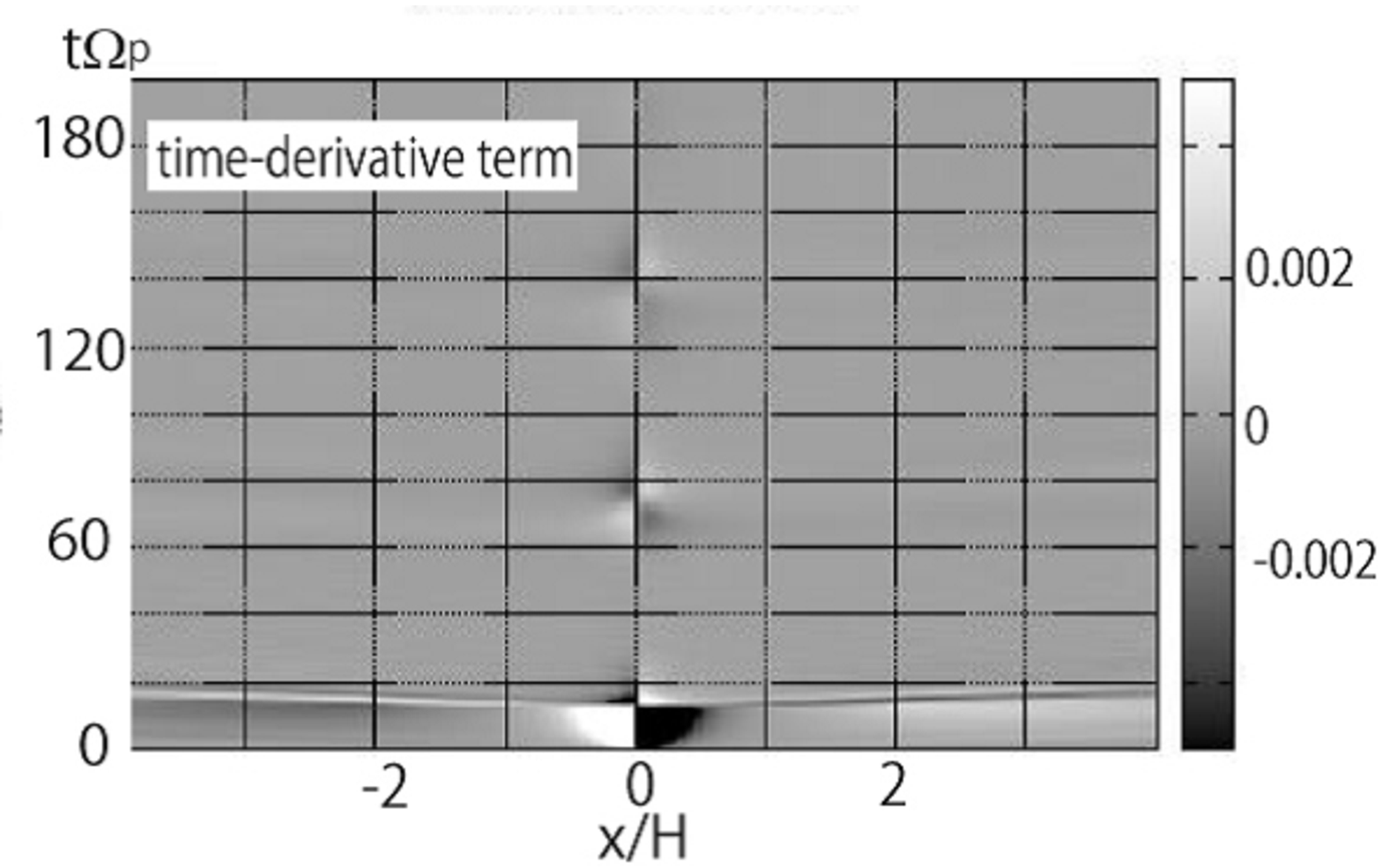}
 \caption{
 The evolution of the source terms of mass flux obtained by the
 calculations with $GM_{\rm p}/Hc^2=0.1$.  Left panel shows $S_v(t,x)$
 given by equation \eqref{source_vort}.  Right panel shows 
 $\partial_t S_t(t,x)$ where $S_t(t,x)$ is given by equation
 \eqref{source_time}.  
 }
 \label{fig:source_evol}
\end{figure}

We further investigate how this mass flux is formed.  
We first look at the source term of the mass flux given by equations
\eqref{source_vort} and \eqref{source_time}.
We have used the results of numerical calculations to calculate the
perturbed values that appear in these two equations.
Figure \ref{fig:source_evol} shows the evolution of the source of mass
flux given by \eqref{source_vort} and \eqref{source_time} for the
calculation with $GM_{\rm p}/Hc^2=0.1$.  We see that the term with 
$\partial_t S(t,x)$ is not significant after the planet mass is fully
introduced at $t\Omega_{\rm p}=12$.  
The term $S_v(t,x)$ depends on time
strongly in the vicinity of the planet, while nearly time-independent
contribution is observed from the region $|x/H|>1.5$.  The source
arising in the vicinity of the planet comes from the vortex making a
horseshoe orbit, while the source at $|x/H|>1.5$ comes from shock
formation.  

\begin{figure}
 \plotone{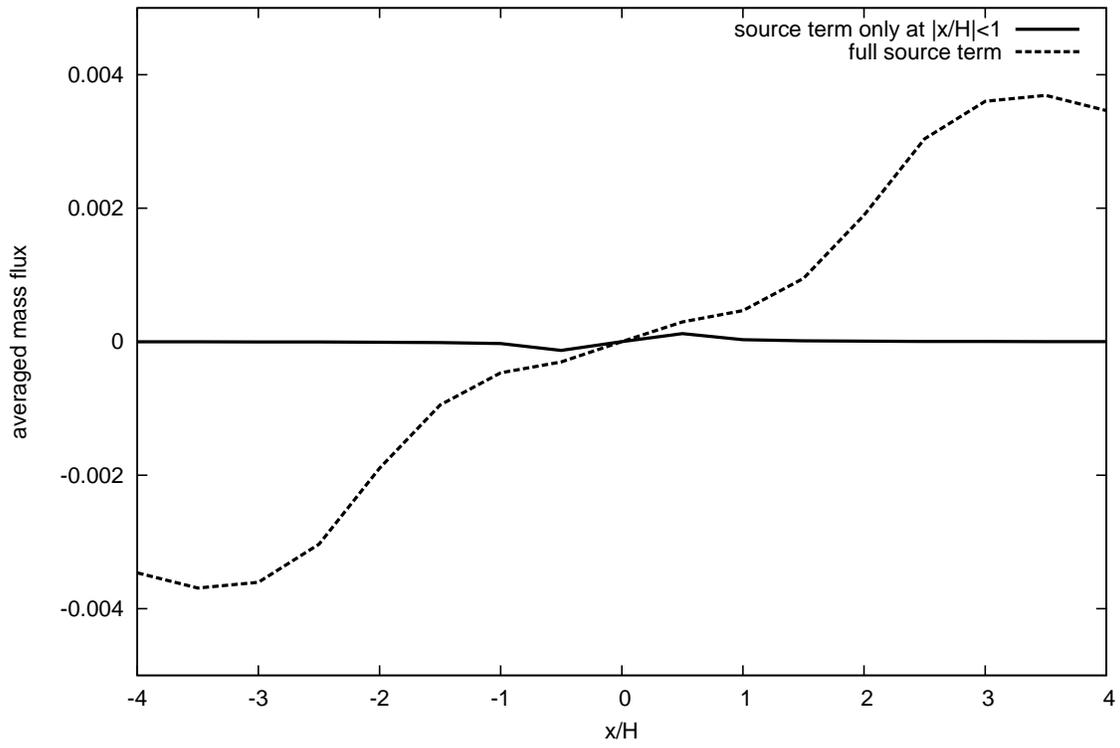}
 \caption{
 Mass flux obtained by second order perturbation theory.  Solid line
 shows the case where the source term is restricted to $|x/H|<1$, while
 the dashed line shows the result with full source term.    
 }
 \label{fig:mdot_sourceterm}
\end{figure}

In order to look at which contribution excites the mass flux
more effectively, we calculate the mass flux by artificially cutting the
source term at $|x/H|>1$.  Figure \ref{fig:mdot_sourceterm} compares the
mass flux obtained by doing so 
and the mass flux obtained by using the full source term. 
We have used equation \eqref{mdot_sol} 
with the source terms obtained by
numerical calculation to derive the mass flux.
It is clear that the source in the vicinity of the planet does not
contribute to the mass flux very much.  Therefore, we conclude that the
mass flux arises from the specific vorticity generated by the shock
dissipation of density wave.  

\begin{figure}
 \plotone{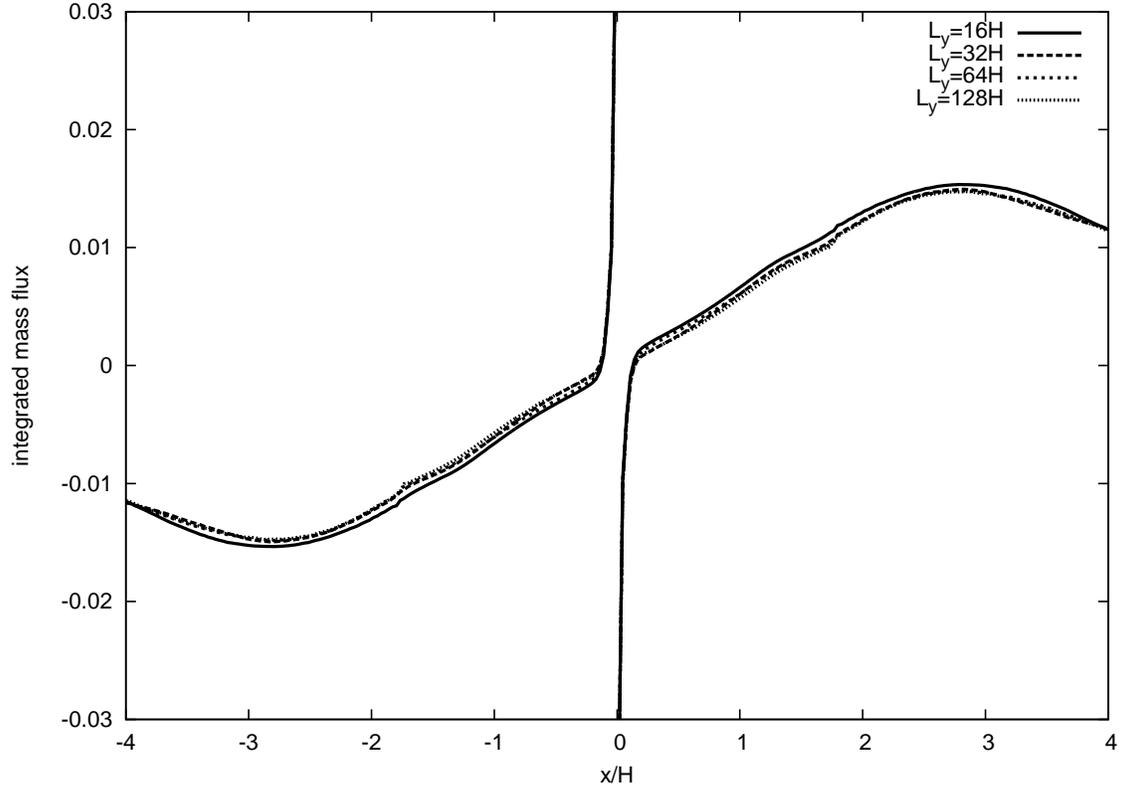}
 \caption{
 Integrated mass flux $\int dy \Sigma v_x$ for box sizes with 
 $L_y/H=16, 32, 64, 128$.  The mass of the planet is 
 $GM_{\rm p}/Hc^2=0.2$ and the data at $t\Omega_{\rm p}=100$ are used.
 } 
 \label{fig:variableLy}
\end{figure}

We now discuss the dependence of mass flux on the box size in the
$y$-direction, $L_y$.
Figure \ref{fig:variableLy} shows the azimuthally integrated mass flux
$\int dy \Sigma v_x $
with $GM_{\rm p}/Hc^2=0.2$ for different box length $L_y$.  
These mass fluxes are directly calculated 
from the numerical simulations.  
The lines show a perfect match for different box sizes.
This can be explained if we notice that the angular momentum flux
carried by the density wave does not depend on the box size.  
As the spiral density wave damps, the angular momentum carried by the
wake is deposited to the background disk to drive the mass flux.  
The total amount of the angular momentum deposited over the
$y$-direction does not depend on the box size, 
since the amount of angular momentum flux carried by the wave does not
depend on the box size.  

We now discuss the timescale for the gap-opening in the vicinity of the
planet.  
As stated at the end of Section \ref{subsec:second}, 
the mass flux in the vicinity of the planet 
is expected to be proportional to $x$, and if
second-order analysis is valid, we expect that the mass flux is
proportional to the square of the planet mass.  We therefore fit the
mass flux within $|x/H|<2$ using the function of the form
\begin{equation}
 \dfrac{1}{L_y} \int dy \Sigma v_x = K \Sigma_0 c 
  \dfrac{x}{H}
  \left( \dfrac{GM_{\rm p}}{Hc^2} \right)^2 
  \left( \dfrac{L_y/H}{32} \right)^{-1},
  \label{massflux_fit}
\end{equation}
where K is a constant that is derived by fitting the results of the
numerical simulations.  We found $K=4.4 \times 10^{-3}$ can fit the
numerical results upto $GM_{\rm p}/Hc^2 \lesssim 0.4$ within 15\% error
for $|x/H|\leq 2$.  Figure \ref{fig:mdot_fit} compares the mass flux
obtained by numerical simulation and the fitting function
\eqref{massflux_fit}.

Since the gap opening timescale may be determined by
\begin{equation}
 \tau_{\rm gap}^{-1} 
  = \dfrac{(1/L_y)\partial_x \int dy \Sigma v_x}{(1/L_y)\int dy \Sigma}, 
  \label{gaptime_num}
\end{equation}
and the denominator is approximated by $\Sigma_0$, 
the gap-opening timescale is proportional to $L_y$.  
Using the result of equation \eqref{massflux_fit}, we have 
\begin{equation}
 \tau_{\rm gap}^{-1} = 4.4\times 10^{-3} \Omega_{\rm p} 
  \left( \dfrac{GM_{\rm p}}{Hc^2} \right)^2
  \left( \dfrac{L_y/H}{32} \right)^{-1}.
  \label{gaptime_est1}
\end{equation}
Note that this timescale applies to the initial phase of gap opening,
and it does not state about how deep the gap will be.  
The box size in the $y$-direction may be interpreted as a circumference
of the disk.  The above expression can be rewritten as
\begin{equation}
 \tau_{\rm gap}^{-1} = 1.1\times 10^{-3} \Omega_{\rm p} 
  \left( \dfrac{GM_{\rm p}}{Hc^2} \right)^2
  \left( \dfrac{H/r_{\rm p}}{0.05} \right)
  \left( \dfrac{2 \pi r_{\rm p}}{L_y} \right).
  \label{gaptime_est2}
\end{equation}

\subsubsection{Instability of Gap Profile}
\label{subsec:gapinst_num}

\begin{figure}
 \plottwo{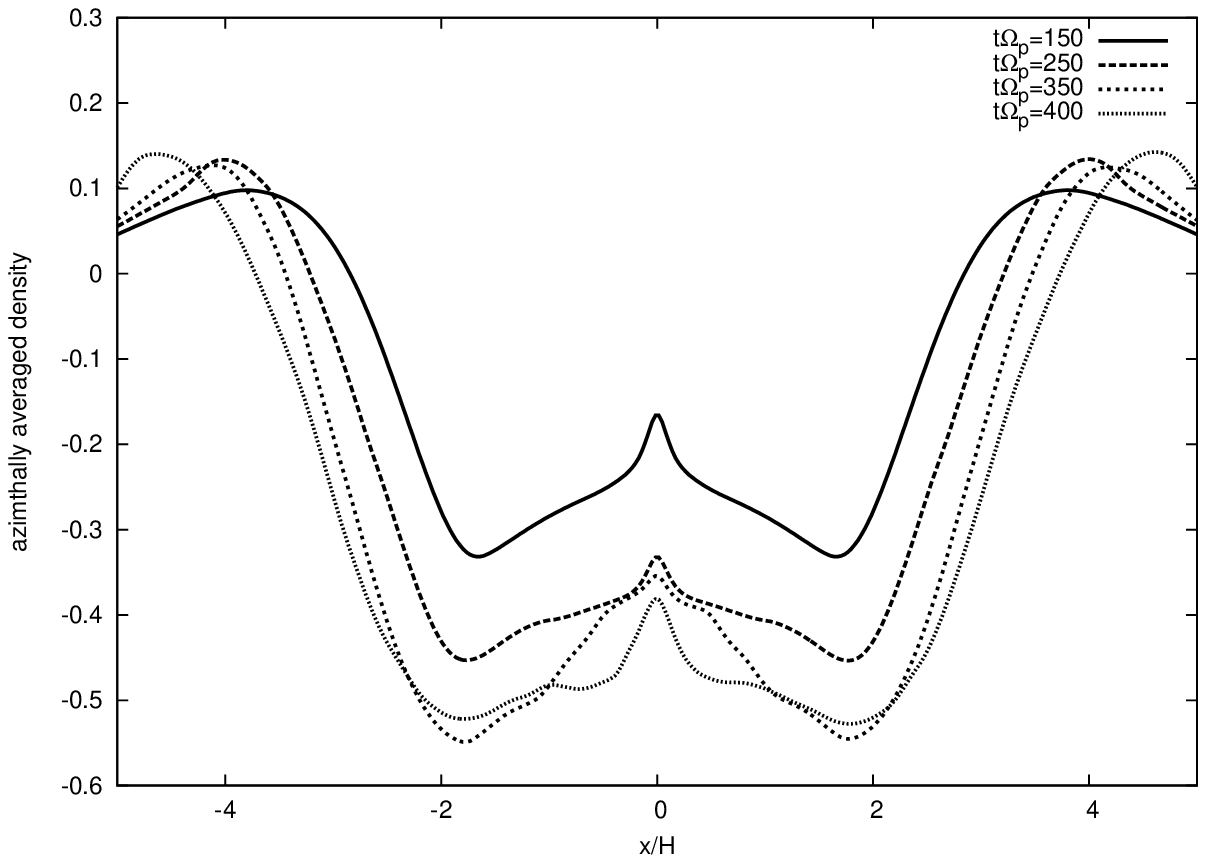}{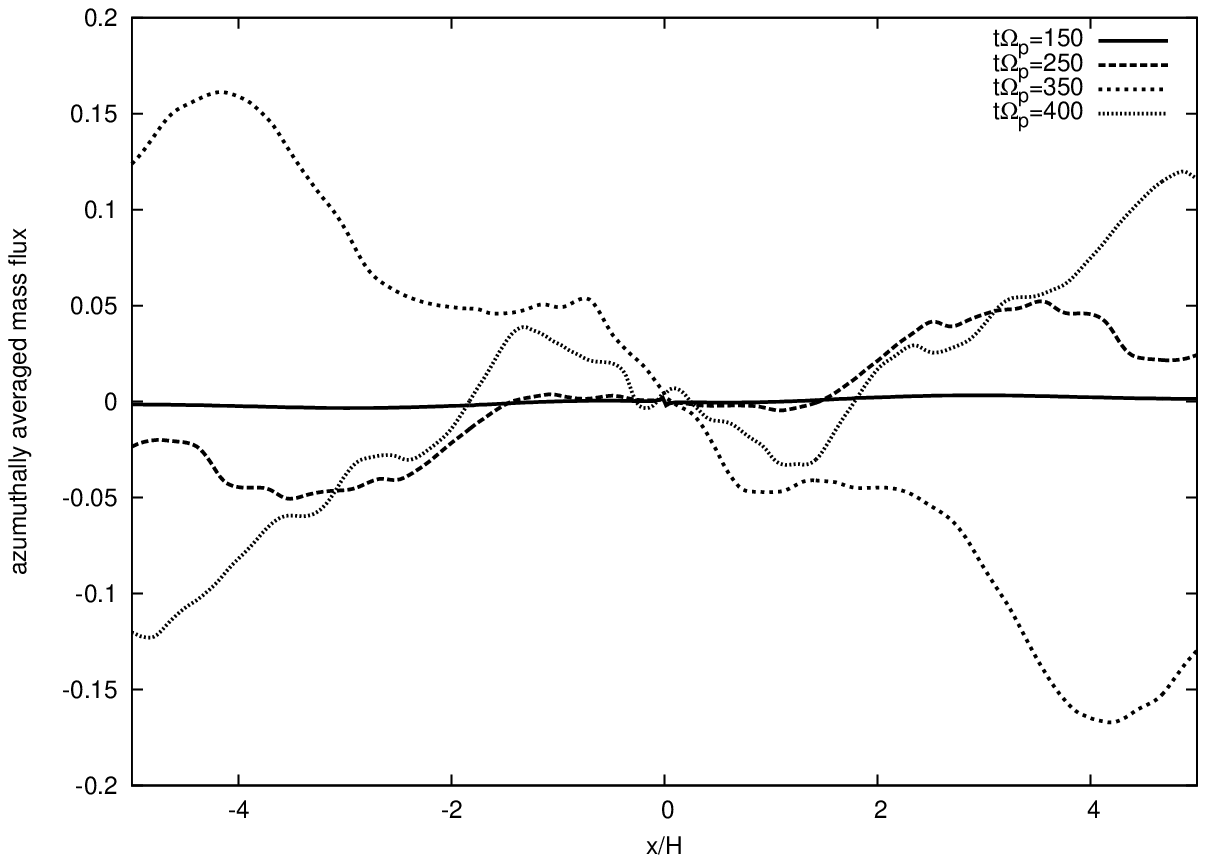}
 \caption{
 Evolution of density profile (left) and the mass flux (right) for the
 calculation with $GM_{\rm p}/Hc^2=0.6$
 } 
 \label{fig:gapinst_M06}
\end{figure}

In Section \ref{subsec:gapinst}, we have seen that the disk with a
density bump can be prone to linear instability.  Figure
\ref{fig:gapinst_M06} shows the evolution of 
azimuthally averaged density perturbation and mass flux in 
the calculation with $GM_p/Hc^2=0.6$.
We see that the mass flux changes the sign very rapidly, 
and the gap depth saturates when
approximately half of the original mass is depleted.  The width of the
gap we have obtained is of the order of several scale height of the
disk.  

It seems that the gap width still increases gradually.
This may be explained quantitatively as follows.  The perturbation by
the planet tries to deposit the angular momentum to the disk in such a
way that the fluid elements to migrate away from the planet.  Turbulence
may try to redistribute the angular momentum, but at the same time,
diffusion in the radial direction occurs because of the turbulence.  
Therefore, not all the angular momentum deposited to the disk is
absorbed by turbulence, and a part of the angular momentum goes to the
background, and the gap gradually becomes wider and wider.  
It may be possible to form a wider gap.  The final state of
the gap may not be able to capture in the local shearing-sheet
calculation since the radial extent of the calculation box is not very
wide.    

Li et al. (2009) obtained a partial gap (approximately half of the gas
depleted) with the width of about $5-7H$
for the parameter $GM_{\rm p}/Hc^2\sim 0.7$ (see Figure 2 of their
paper).    
Although 
it may be possible that the final width and the shape of the
gap is not well-determined in the shearing-sheet calculations, the gap
formation process described 
in the previous sections may be valid.  We interpret the gap depth
and width obtained in our calculation as minimum values.   
Planets may potentially be able to open a deeper and wider gap.
We note that Crida et al. (2006) and de Val-Borro et al. (2007) obtained
deeper and wider gap for a Jupiter mass planet, although they assumed
very low or zero viscosity. 

\section{Discussion: Gap Formation in a Protoplanetary Disk}
\label{sec:gapform}

In this section, using the results of numerical calculations, we discuss
the validity of conventional one-dimensional model for gap formation and
the condition of gap opening.  

\subsection{One-Dimensional Model}
\label{subsec:1d}

The gap formation processes are modeled using one-dimensional disk
evolution model.  Thommes et al. (2008) uses a model in which  
the torque exerted at the Lindblad resonances are 
directly deposited to the disk to change the semi-major axis of the
fluid particle.  Rafikov (2002b) 
uses a one-dimensional model based on Lynden-Bell and 
Pringle (1974) type surface density evolution
equation to model gap formation processes. 
We investigate whether such treatment can reproduce the evolution of the
density profile obtained in two-dimensional calculations.

Although we have performed shearing-sheet analyses, we first note that
global calculation and local shearing-sheet calculation share very
similar property if we consider
the azimuthally-averaged one-dimensional model.  
This can be investigated by comparing the one-dimensional model derived
using global model and local model (see Appendix \ref{app:1Dmodel}).

\begin{figure}
 \plotone{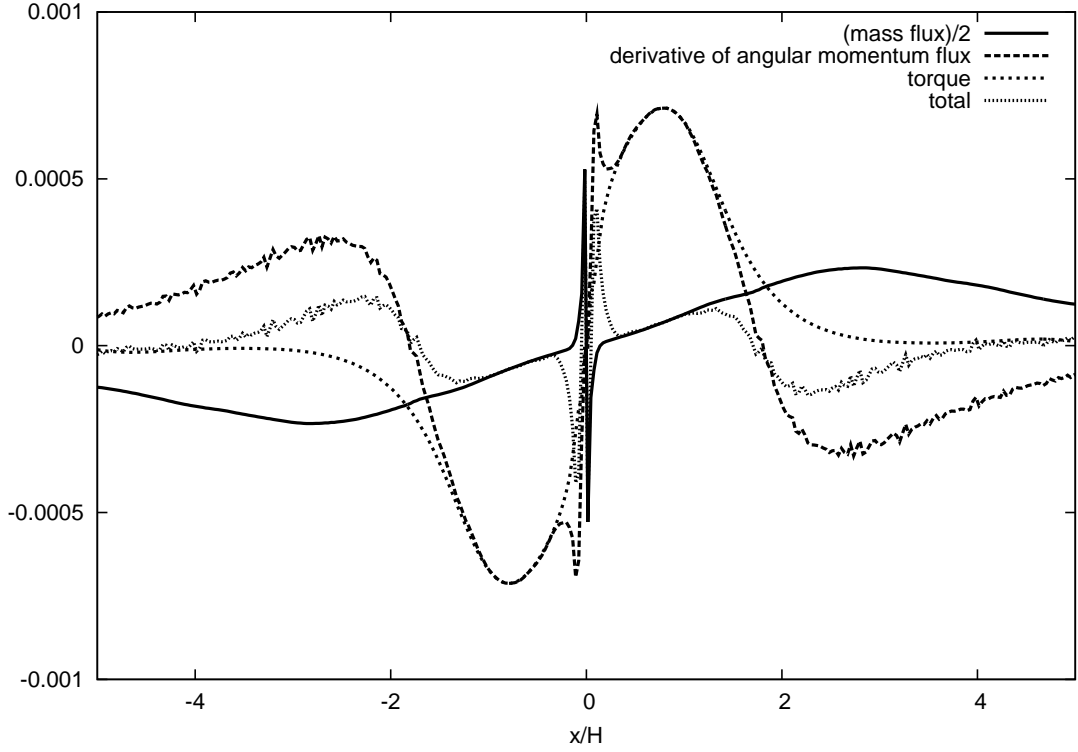}
 \caption{
 Comparison of each term in equation \eqref{mdot_angmom_1D_local}.  
 Data with $GM_{\rm p}/Hc^2=0.2$ at $t\Omega_{\rm p}=100$ is used.  The
 line ``(mass flux)/2'' is the left hand side of equation
 \eqref{mdot_angmom_1D_local}, ``derivative of angular momentum flux''
 is the first term of the right hand side (sign inverted), and
 ``torque'' is the second term of the right hand side.  The line
 ``total'' is the value when all the terms are taken to the left hand
 side, and should be zero if equation \eqref{mdot_angmom_1D_local} is
 strictly satisfied. 
 } 
 \label{fig:mdot_angmom_relation}
\end{figure}

Exact one-dimensional evolution model is given by equations
\eqref{EoC_1Dmodel_local} and \eqref{EoMy_1Dmodel_local_mod} in 
local approximation.  Whether the evolution model 
\eqref{dens_evolution_1D_local} can be used depends on whether the 
approximation made in deriving this equation is appropriate.  This can
be seen whether mass flux, angular momentum, and the torque exerted by
the planet are related as equation \eqref{mdot_angmom_1D_local}.  
In Figure \ref{fig:mdot_angmom_relation}, we compare each term of
equation \eqref{mdot_angmom_1D_local}.  We observe that the relationship
of this equation is not satisfied, indicating the time evolution of
$\Sigma \delta v_y$ occurs.  
We have checked that the exact time evolution 
equation \eqref{EoMy_1Dmodel_local_mod} is satisfied.  

We therefore argue that time evolution of density and rotation
profile is given by equation \eqref{EoC_1Dmodel_local} and
\eqref{EoMy_1Dmodel_local_mod}, and in constructing one-dimensional
model, it is necessary to specify the model of the mass flux, as well as
the torque and angular momentum flux.  It is especially important in
considering the gap formation in the vicinity of the planet orbit.  For
example, Rafikov (2002b) predicts nothing happens in the vicinity of the
planet in an inviscid model unless the spiral density wave shocks.  In
two-dimensional simulation, in contrast, 
we also observe gap opening even in low-mass planet calculations.  
As the gap opens, rotation velocity of the gas is also modified in order
to balance the pressure gradient, as Crida et al. (2006) pointed out.

\subsection{Gap Opening Condition in an Inviscid Disk}
\label{subsec:gapcriterion}

We have seen that low-mass planets can 
potentially open a gap, or at least a ring with low surface density, in
an inviscid disk.  We now discuss the minimum mass of the planet
that can open the gap.  
Low mass planets are prone to type I migration and therefore, if the
planet migrates before it opens a gap, gap formation is impossible.
Since the width of the gap is the order of scale height $H$, we compare
the timescale of the planet migration over length $H$ and the gap
formation timescale.  The planet migration timescale in an isothermal
disk is estimated by Tanaka et al. (2001) as
\begin{equation}
 \tau_{\rm mig}^{-1} \sim 3\Omega_{\rm p} 
  \dfrac{M_{\rm p}}{M_{\ast}} \dfrac{\Sigma r_{\rm p}^2}{M_{\ast}}
  \left( \dfrac{r_{\rm p}}{H} \right)^3.
\end{equation}
Note that the power of $H/r_{\rm p}$ is $3$ because we consider the
migration over the radial length $H$.  For gap formation timescale, we
use the 
results of numerical calculations given in Section \ref{subsec:result}.
The gap opening timescale $\tau_{\rm gap}^{-1}$ is given by equation
\eqref{gaptime_est1}.  
If $\tau_{\rm mig}^{-1} < \tau_{\rm gap}^{-1}$, we expect the gap will
open in the vicinity of the planet.  
Comparing $\tau_{\rm mig}$ and $\tau_{\rm gap}$, we obtain
\begin{equation}
 \dfrac{M_{\rm p}}{M_{\ast}} \gtrsim  2.0 \times 10^{-5}
  \left( \dfrac{H/r_{\rm p}}{0.05} \right)^3
  \left( \dfrac{L_y/H}{32} \right)
  \dfrac{\Sigma}{2\times 10^3 \mathrm{gcm}^{-3}}
  \left( \dfrac{r_{\rm p}}{1\mathrm{AU}} \right)^2
  \left( \dfrac{M_{\ast}}{M_{\odot}} \right)^{-1}.
  \label{cond_inviscid}
\end{equation}
We have used typical values of protoplanetary disk at $1\mathrm{AU}$ to
estimate the number.  
We note that the gap-opening mass scales with the box size $L_y$.  If we
assume $L_y=2\pi r_{\rm p}$ and $H/r_{\rm p}=0.05$, $L_y/H = 125.66$,
which gives $M_{\rm p}/M_{\ast} \gtrsim 8 \times 10^{-5}$, which is
below the criterion given in previous studies, e.g., equation
\eqref{cond_crida}.  
Equation \eqref{cond_inviscid} rewritten by using $2\pi r_{\rm p}$ reads 
(see also equation\eqref{gaptime_est2})
\begin{equation}
 \dfrac{M_{\rm p}}{M_{\ast}} \gtrsim  8.0 \times 10^{-5}
  \left( \dfrac{H/r_{\rm p}}{0.05} \right)^2
  \left( \dfrac{L_y}{2\pi r_{\rm p}} \right)
  \dfrac{\Sigma}{2\times 10^3 \mathrm{gcm}^{-3}}
  \left( \dfrac{r_{\rm p}}{1\mathrm{AU}} \right)^2
  \left( \dfrac{M_{\ast}}{M_{\odot}} \right)^{-1}.
  \label{cond_inviscid2}
\end{equation}

It is interesting to compare equation \eqref{cond_inviscid} with the
previous criterion \eqref{cond_crida}.  Equation 
\eqref{cond_inviscid} is derived by comparing the gap opening timescale
and type I migration timescale, and this is qualitatively different from
the way the previous criterion is derived.  For example, 
the gap opening criterion
derived in this study, equation \eqref{cond_inviscid}, depends on the
disk mass, since type I migration timescale scales with the disk mass.
We also note that the type I migration timescale may be much longer if
the corotation torque and non-barotropic effects 
are considered (Paardekooper et al. 2010ab).  In this case, our results
indicate that the gap may be opened for lower mass planets, although
more thorough analyses of non-barotropic effects on 
spiral density wave and mass flux are necessary.  

If the disk is in turbulent state, the effective viscosity can also act
to fill the gap.  If we use standard $\alpha$-prescription for turbulent
viscosity 
$\nu = \alpha c H$,
the timescale of viscous diffusion over length scale $\sim H$ is
\begin{equation}
 \tau_{\rm vis}^{-1} \sim \alpha \Omega_{\rm p}.
\end{equation}
If $\tau_{\rm vis}$ is shorter than $\tau_{\rm gap}$, we expect
that the gap is filled.  This leads to another condition,
\begin{equation}
 \alpha \lesssim 1.1 \times 10^{-4} 
  \left( \dfrac{H/r_{\rm p}}{0.05} \right)^{-6}
  \left( \dfrac{M_{\rm p}/M_{\ast}}{2 \times 10^{-5}} \right)^2
  \left( \dfrac{L_y/H}{32} \right)^{-1},
\label{gapcond_vis}
\end{equation}
or if we use $2\pi r_{\rm p}$ instead of $L_y$, 
\begin{equation}
 \alpha \lesssim 2.8 \times 10^{-5} 
  \left( \dfrac{H/r_{\rm p}}{0.05} \right)^{-5}
  \left( \dfrac{M_{\rm p}/M_{\ast}}{2 \times 10^{-5}} \right)^2
  \left( \dfrac{2\pi r_{\rm p}}{L_y} \right).
\label{gapcond_vis2}
\end{equation}
The disk must be quiet in order for a low-mass planets to open up the
gap, but we note that the condition is very sensitive to the disk aspect
ratio.  
We note that the previous results for 
gap-opening criterion derived by Crida et
al. (2006) is somewhat different from equation \eqref{gapcond_vis}.
Since we derive equation \eqref{gapcond_vis} using a crude
order-of-magnitude estimate for viscous diffusion, it is necessary to
perform more systematic parameter study in order to investigate how much
viscosity is necessary to halt gap-opening by a planet.

The gap-opening criteria, equations \eqref{cond_inviscid} and
\eqref{gapcond_vis}, are more exactly the conditions for the ``formation
of under-dense region around the planetary orbit''.  We compare the
timescale for the decrease of surface density, which is derived from the
mass flux induced by the planet, and the type I migration timescale or
viscous diffusion timescale.  
Therefore, these conditions concern the initial phase of the
gap-opening, and 
the timescale argument does not concern 
how deep the gap will be as a result of long-term evolution.  
The condition by Crida et al. (2006) is, on the other hand, the 
condition for the formation of ``deep gap'', 
or more quantitatively the formation of the gap where $90\%$ of the
original amount of gas is depleted.  
Therefore, the conditions
are not necessarily in contradiction with each other.  
For example, in the case of the numerical calculation with 
$GM_{\rm p}/Hc^2=0.6$, $3H/4r_H \sim 1.3$, which is slightly above the
gap-opening criterion by Crida et al. (2006).  
In figure \ref{fig:gapinst_M06}, we observe a gap where approximately 
$50-60\%$ of the original gas mass is depleted, which is just below the
depth of the gap which is predicted by Crida et al. (2006)  
This may suggest that the prediction by Crida et al. (2006) and our
calculation is in qualitative agreement, although the setting of
our calculation is different from theirs.    
Our condition,
however, still has a new aspect since it states that the gap-opening
process may be different in case of inviscid case.  Non-linear evolution
of density wave and the feedback onto the disk can be important when we
consider an inviscid disk.

\section{Summary and Future Prospects}
\label{sec:summary}

In this paper, we have investigated the disk-planet interaction and 
 the evolution of surface density profile of the disk 
using analytic methods and high-resolution numerical
calculations within the framework of shearing-sheet approximation.   
We have shown some analytic framework to understand the non-linearity of 
disk-planet interaction.  Formation of specific vorticity 
by shock dissipation of density wave can be a source of disk mass flux
in the vicinity of the planet, which results in the
formation of gap around the embedded planet.  
We estimate the conditions of 
 the formation of the underdense region, or gap formation, 
 around the planetary orbit, 
and it is indicated that 
 a planet with 20-30 Earth mass at $1\mathrm{AU}$ 
will open a gap in a standard Minimum Mass Solar Nebula with 
$H/r\sim0.05$. 
We note that our condition applies 
to the initial phase of the gap opening, and  
the condition depends on the type I migration timescale, which
can be much longer 
than that given by Tanaka et al. (2002) formula.  
If type I migration is halted, much lower mass
planets can open a gap.  
In an inviscid case, which we have explored in this paper, the width of
the gap is of the order of the disk scale height.  
We note that we have not discussed the final gap depth in detail in this
paper, although it is indicated that the gap opening process ends by the
onset of hydrodynamic instability when approximately half of the mass is
depleted.  One may call such 
a shallow underdense region a ``dip'' or ``partial gap'', 
rather than gap.  
  We have also discussed that 
classical one-dimensional model of disk evolution fails to describe gap
opening process, and more
refined one-dimensional is necessary to explain gap opening.  

We have focused on the two-dimensional local shearing-sheet
calculations.  We have discussed that the final width of the gap may be
larger than we have obtained in this calculation.  The gap depth and
width we have obtained may be interpreted as minimum values.  Global
calculations with high resolution are necessary to construct more
quantitative model for gap formation.  
We also note that the
discussion using the specific vorticity we have mentioned 
a number of times in this paper 
may be appropriate only in the two-dimensional
calculations.  Although we expect that the two-dimensional model is
adequate for the investigation of spiral density wave qualitatively, 
three-dimensional calculations might result in the quantitatively
different condition for gap opening.

\acknowledgments
Authors thank the referee for useful comments that improved the paper.
This work was supported by the Grant-in-Aid for the Global COE Program
``The Next Generation of Physics, Spun from Universality and Emergence''
from the Ministry of Education, Culture, Sports, Science and Technology
(MEXT) of Japan.
The numerical calculations were in part 
carried out on Altix3700 BX2 at YITP in Kyoto University. 
Data analyses were in part carried out on the data analysis server
at Center for Computational Astrophysics, CfCA, of National Astronomical
Observatory of Japan.
The page charge of this paper is 
supported by CfCA. 
T. M. is supported by Grants-in-Aid (22$\cdot$2942) from MEXT of Japan.
S. I. is
supported by Grants-in-Aid (15740118, 16077202, and 18540238) from MEXT
of Japan.
T. K. S. is supported by Grants-in-Aid (20740100) from MEXT of Japan.  

\appendix

\section{Derivation of One-Dimensional Model}
\label{app:1Dmodel}

In this section, we compare one-dimensional dynamical evolution model
derived from global and local model, and show that they share very
similar properties.

\subsection{Global Model}

We use a cylindrical coordinate system $(r,\phi)$.  In an inertial
frame, a set of fluid equations in two-dimensional global model is given
by 
\begin{equation}
 \dfrac{\partial \Sigma}{\partial t} +
  \dfrac{1}{r}\dfrac{\partial}{\partial r} 
  \left( r\Sigma v_r \right) 
  + \dfrac{1}{r} \dfrac{\partial}{\partial \phi} 
  \left( \Sigma v_{\phi} \right) = 0,
  \label{EoC_global}
\end{equation}
\begin{equation}
 \dfrac{\partial v_r}{\partial t} + v_r \dfrac{\partial v_r}{\partial r} 
  + \dfrac{v_{\phi}}{r} \dfrac{\partial v_r}{\partial \phi} 
  - \dfrac{v_{\phi}^2}{r} =
  - \dfrac{1}{\Sigma} \dfrac{\partial p}{\partial r} 
  - \dfrac{d \Psi_{\ast}}{dr} 
  - \dfrac{\partial \psi_{\rm p}}{\partial r} ,
  \label{EoMx_global}
\end{equation}
\begin{equation}
 \dfrac{\partial v_{\phi}}{\partial t} 
  + v_r \dfrac{\partial v_{\phi}}{\partial r} 
  + \dfrac{v_{\phi}}{r} \dfrac{\partial v_r}{\partial \phi} 
  - \dfrac{v_r v_{\phi}}{r} =
  - \dfrac{1}{\Sigma r} \dfrac{\partial p}{\partial \phi} 
  - \dfrac{1}{r}\dfrac{\partial \psi_{\rm p}}{\partial \phi} ,
  \label{EoMy_global}
\end{equation}
where $\Psi_{\ast}$ is the gravitational potential of the central star
which is assumed to be axisymmetric.  
The rotation profile of the background disk is given by
\begin{equation}
 r\Omega^2(r) = \dfrac{d\Psi_{\ast}}{dr} .
\end{equation}

We now derive a one-dimensional model for disk evolution by taking the
azimuthal average of  
equation of continuity \eqref{EoC_global} and conservation of angular
momentum, which is essentially equation \eqref{EoMy_global}.
We decompose the azimuthal velocity into background part and
perturbation, 
\begin{equation}
 v_{\rm \phi} = r\Omega(r) + \delta v_{\phi}
  \label{vphi_decompose}
\end{equation}
but we do not use linear approximation.

Azimuthally averaged (denoted by bar) equation of continuity is
\begin{equation}
 \dfrac{\partial\overline{\Sigma}}{\partial t} +
  \dfrac{1}{r}\dfrac{\partial}{\partial r} 
  \left( r \overline{\Sigma v_r} \right) = 0,
  \label{1D_global_EoC}
\end{equation}
and the azimuthally averaged angular momentum conservation is
\begin{equation}
 \dfrac{\partial}{\partial t} 
  \left( r \overline{\Sigma v_\phi} \right)
 + \dfrac{1}{r} \dfrac{\partial}{\partial r}
 \left( r^3\Omega \overline{\Sigma v_r }  + 
 r^2 \overline{\Sigma v_r \delta v_{\phi}} \right) 
 = T(r,t), 
  \label{1D_global_EoMy}
\end{equation}
where
\begin{equation}
 T(r,t) = - \overline{\Sigma \partial_{\phi} \psi_{\rm p}}
\end{equation}
is the torque exerted by the planet.

Decomposing $v_{\phi}$ as equation \eqref{vphi_decompose}, 
equation \eqref{1D_global_EoMy} can be further rewritten, with the aid
of equation \eqref{1D_global_EoC}, 
\begin{equation}
 \dfrac{\partial}{\partial t} 
  \left( r \overline{\Sigma \delta v_\phi} \right)
 + \dfrac{1}{r} \dfrac{\partial}{\partial r}
 \left( r^2 \overline{\Sigma v_r \delta v_{\phi}} \right) 
 = -(r^2\Omega)^{\prime} T(r,t), 
  \label{1D_global_EoMy_mod}
\end{equation}
where $\prime$ denotes the derivative with respect to $r$.

Upto equation \eqref{1D_global_EoMy_mod}, 
our treatment is exact.  In the standard one-dimensional
model, approximation $r\Omega \gg \overline{\delta v_r}$ is used to
neglect the time-derivative of equation \eqref{1D_global_EoMy_mod} to
obtain the relationship between the mass flux and angular momentum flux
(Balbus and Papaloizou 1999).  This approximation means that one
neglects the evolution of the rotation profile.  We then obtain the
relation between mass flux, angular momentum flux, and torque, 
\begin{equation}
 \left(r^2 \Omega\right)^{\prime} \overline{\Sigma v_r}
  = - \dfrac{1}{r} \dfrac{\partial}{\partial r}
  \left( \Sigma v_r \delta v_{\phi} \right)
  + T(r,t),
\end{equation}
and the evolution of surface density is obtained from equation
\eqref{1D_global_EoC}
\begin{equation}
 \dfrac{\partial \overline{\Sigma}}{\partial t}
  - \dfrac{1}{r} \dfrac{\partial}{\partial r} 
  \left[ \dfrac{1}{(r^2\Omega)^{\prime}} 
   \dfrac{\partial}{\partial r} 
   \left( r^2 \overline{\Sigma v_r \delta v_{\phi}}
   \right)
  \right] =
  - \dfrac{1}{r} \dfrac{\partial}{\partial r} 
  \left[ \dfrac{r T(r,t)}{(r^2\Omega)^{\prime}}
  \right] .
  \label{dens_evolution_1D}
\end{equation}
If $\alpha$-prescription is used for the angular momentum flux and the
planet is absent, this is the model by Lynden-Bell and Pringle (1974).
Rafikov (2002b) used this equation to study gap formation.

\subsection{Local Model}

It is possible to derive equations analogous to global model in
averaging over the $y$-direction in the shearing-sheet approximation.
Taking the $y$-average of equation of continuity, we obtain
\begin{equation}
 \dfrac{\partial \overline{\Sigma}}{\partial t} 
  + \dfrac{\partial}{\partial x} \overline{\Sigma v_x} = 0
  \label{EoC_1Dmodel_local}
\end{equation}
and from the equation of motion in the $y$-direction in a conservation
form,
\begin{equation}
 \dfrac{\partial}{\partial t} \overline{\Sigma v_y}
  -\dfrac{2}{3}\Omega_{\rm p} x \dfrac{\partial}{\partial x}
  \overline{\Sigma v_x} 
  + \dfrac{\partial}{\partial x} 
  \overline{\Sigma v_x \delta v_y} 
  = -2\Omega_{\rm p} \overline{\Sigma v_x} 
  + T_{\rm loc} (t,x),
  \label{EoMy_1Dmodel_local}
\end{equation}
where
\begin{equation}
 T_{\rm loc} = -\overline{\Sigma \partial_y \psi_{\rm p}}.
\end{equation}
Decomposing the first term in equation \eqref{EoMy_1Dmodel_local} into
the background and perturbation, we obtain
\begin{equation}
 \dfrac{\partial}{\partial t} \overline{\Sigma \delta v_y}
    + \dfrac{\partial}{\partial x} 
  \overline{\Sigma v_x \delta v_y} 
  = -\dfrac{1}{2}\Omega_{\rm p} \overline{\Sigma v_x} 
  + T_{\rm loc} (x).
  \label{EoMy_1Dmodel_local_mod}
\end{equation}
This corresponds to equation \eqref{1D_global_EoMy_mod} in the global
model.  

If we approximate that 
$ |(3/2)\Omega_{\rm p} x| \gg \overline{\delta v_y}$, 
mass flux and angular momentum flux are related by
\begin{equation}
  \dfrac{1}{2}\Omega_{\rm p} \overline{\Sigma v_x} = 
   - \dfrac{\partial}{\partial x} 
  \overline{\Sigma v_x \delta v_y} 
   + T_{\rm loc} (x).
  \label{mdot_angmom_1D_local}
\end{equation}
we obtain
\begin{equation}
 \dfrac{\partial \overline{\Sigma}}{\partial t} + 
  \dfrac{\partial}{\partial x} 
  \left[  
   -\dfrac{2}{\Omega_{\rm p}} \dfrac{\partial}{\partial x} 
   \overline{\Sigma v_x \delta v_y}
  \right] =
  -\dfrac{2}{\Omega_{\rm p}} T_{\rm loc}(x),
  \label{dens_evolution_1D_local}
\end{equation}
which corresponds to equation \eqref{dens_evolution_1D} in global
model.  

We also note that the equation for mechanical energy loss is also
analogous in global and local model.

\section{Linear Stability Analysis of a Disk with a Gap}
\label{app:RWI}

In this section, 
we show the outline of the linear stability analysis of a disk when the
surface density structure is not uniform.   

We consider a disk without a planet for simplicity.
The equations we consider are equations \eqref{EoC_full}
and \eqref{EoM_full} without $\psi_{\rm p}$.
We assume that the background disk is axisymmetric with density profile
$\Sigma_0(x)$.  We denote the background values with subscript ``0''.    
In the background state, the pressure
gradient must be balanced by Coriolis force and therefore,
\begin{equation}
 v_{y,0}(x) = -\dfrac{3}{2}\Omega_{\rm p} x 
  + \dfrac{c^2}{2\Omega_{\rm p}} \dfrac{1}{\Sigma_0}\dfrac{d\Sigma_0}{dx}
  \equiv U(x),
  \label{U_def}
\end{equation}
and 
\begin{equation}
 v_{x,0}=0.
\end{equation}

We now consider linear perturbation.  Perturbed values are denoted by 
$\delta$ and consider the solution proportional to 
$\exp[-i\omega t + ik_y y]$.
Linear perturbation is then
\begin{equation}
 -i \tilde{\omega} \dfrac{\delta \Sigma}{\Sigma_0} 
 + \dfrac{1}{\Sigma_0} \dfrac{d \Sigma_0}{dx} \delta v_x + 
 \dfrac{d}{d x} \delta v_x + ik_y \delta v_y = 0,
\end{equation}
\begin{equation}
 -i \tilde{\omega} \delta v_x 
 + c^2 \dfrac{d}{d x} \dfrac{\delta \Sigma}{\Sigma_0} 
 - 2\Omega_{\rm p} \delta v_y = 0,
\end{equation}
\begin{equation}
 -i \tilde{\omega} \delta v_y 
 + c^2 ik_y \dfrac{\delta \Sigma}{\Sigma_0} + 
 \left( 2\Omega_{\rm p} + \dfrac{dU}{dx} \right)\delta v_x = 0,
\end{equation}
where $\tilde{\omega}(x) \equiv \omega - k_y U(x)$.
From these equations, we can derive a single second-order ordinary
differential equation for $\Phi \equiv \Sigma_0 \delta v_x$,
\begin{equation}
 \dfrac{d}{dx} \left[
		    \dfrac{1}{\Sigma_0 (\tilde{\omega}^2 - c^2 k_y^2)} 
	\dfrac{d\Phi}{dx}  \right] +
 \left[ \dfrac{k_y}{2\Omega_{\rm p}\tilde{\omega}}
 \left( \dfrac{\kappa^2}{\Sigma_0 (\tilde{\omega}^2 
  - c^2 k_y^2)} \right)^{\prime} + \dfrac{\tilde{\omega}^2 - c^2 k_y^2 -
 \kappa^2}{c^2 \Sigma_0 (\tilde{\omega}^2 - c^2 k_y^2)} \right] \Phi = 0, 
\label{gapinst_ode}
\end{equation}
where
\begin{equation}
 \kappa^2(x) = 2\Omega_{\rm p}(2\Omega_{\rm p} + U^{\prime}(x))
\end{equation}
and $\prime$ denotes the derivative with respect to $x$.
Equation \eqref{gapinst_ode} with an appropriate boundary condition
constructs an eigenvalue problem for $\omega$.
Since full analyses of equation \eqref{gapinst_ode} is not the scope
of this paper, we briefly outline the qualitative results about the
instability of gap profile that can be derived from equation
\eqref{gapinst_ode}.  

For an axisymmetric mode ($k_y=0$), equation \eqref{gapinst_ode} becomes 
\begin{equation}
 \dfrac{d}{dx} \left[ \dfrac{1}{\Sigma_0} \dfrac{d\Phi}{dx} \right]
  + \dfrac{\omega^2 - \kappa^2}{c^2 \Sigma_0} \Phi = 0.
  \label{gapinst_axisym}
\end{equation}
We multiply $\Phi^{\ast}$ to this equation and integrate over $x$.
Assuming that the perturbation vanishes at $|x|\to\infty$ and
integrating by part, we obtain
\begin{equation}
  \int dx \dfrac{1}{\Sigma_0} \left|\dfrac{d\Phi}{dx}\right|^2
   + \int dx \dfrac{\kappa^2}{c^2 \Sigma_0} \left| \Phi \right|^2
   = \int dx \dfrac{\omega^2}{c^2 \Sigma_0} \left| \Phi \right|^2.
\end{equation}
We first note that from this equation, 
the eigenvalue $\omega^2$ must be real.  For
instability, $\omega^2$ must be negative and therefore, $\kappa^2$ must
be negative at some point in $x$.  This necessary condition for
instability is the well-known Rayleigh criterion.

We can further proceed by using the independent variable
\begin{equation}
 g(x) = \dfrac{1}{\sqrt{\Sigma_0}} \Phi.
\end{equation}
Equation \eqref{gapinst_axisym} now becomes
\begin{equation}
 \dfrac{d^2 g}{dx^2} + \left[ \dfrac{\omega^2}{c^2} 
			- \left( \dfrac{\Omega_{\rm p}^2}{c^2} +
			   \dfrac{1}{\sqrt{\Sigma_0}}\dfrac{d^2}{dx^2}
			  \sqrt{\Sigma_0} \right) \right] g
 = 0
  \label{fulleqn_sch}
\end{equation}
Defining
\begin{equation}
 E = \dfrac{\omega^2}{c^2}
\end{equation}
and
\begin{equation}
 V(x) = \dfrac{\Omega_{\rm p}^2}{c^2} 
  + \dfrac{1}{\sqrt{\Sigma_0}}\dfrac{d^2}{dx^2}\sqrt{\Sigma_0},
\end{equation}
equation \eqref{fulleqn_sch} can be looked as a Scr\"{o}dinger equation with
energy $E$ and potential $V(x)$,  
\begin{equation}
 \dfrac{d^2 g}{dx^2} + [E-V(x)]g(x) = 0.
\end{equation}
If there is a ``bound state'' with energy eigenvalue $E<0$, the system
is unstable.  Necessary condition for instability is therefore that there
exists a point where $V(x)<0$, or
\begin{equation}
 \dfrac{1}{\Sigma_0^{1/2}}\dfrac{d^2}{dx^2}\Sigma_0^{1/2} < -
  \dfrac{\Omega_{\rm p}^2}{c^2}.
\end{equation}   
Therefore, if there exists a sharp pressure bump, 
the system is unstable against axisymmetric perturbation.  

For a non-axisymmetric mode ($k_y\neq 0$), the analysis becomes more
complex.  However, we can make a progress when we assume that there is a
mode confined close to corotation, $\tilde{\omega}=0$.  Assuming
$\tilde{\omega} \sim 0$ so $\tilde{\omega}-c^2 k_y^2 \sim - c^2 k_y^2$, 
equation \eqref{gapinst_ode} becomes
\begin{equation}
 \dfrac{d}{dx} \left[ \dfrac{1}{\Sigma_0} \dfrac{d\Phi}{dx}  \right]
  - \left[ \dfrac{c^2 k_y^2 + \kappa^2}{c^2 \Sigma_0} 
     - \dfrac{1}{\tilde{\omega}}\dfrac{k_y}{2\Omega_{\rm p}} \left(
							      \dfrac{\kappa^2}{\Sigma_0}
							     \right)^{\prime}
    \right] \Phi = 0
\end{equation}
Multiplying $\Phi^{\ast}$ to this equation and integrate over $x$, we
obtain
\begin{equation}
 \int dx \left[ \dfrac{1}{\Sigma_0}\left| \dfrac{d\Phi}{dx} \right|^2
	 + \dfrac{c^2 k_y^2 + \kappa^2}{c^2\Sigma_0}\left| \Phi
						    \right|^2 \right]
 = \int dx \dfrac{1}{\tilde{\omega}} \dfrac{k_y}{2\Omega_{\rm p}}
 \left(\dfrac{\kappa^2}{\Sigma_0} \right)^{\prime} \left| \Phi \right|^2 
\end{equation}
Taking the imaginary part of this equation, we have
\begin{equation}
 \mathrm{Im}(\omega) \int dx \dfrac{1}{\left|\tilde{\omega}\right|^2}
  \dfrac{k_y}{2\Omega_{\rm p}}
 \left(\dfrac{\kappa^2}{\Sigma_0} \right)^{\prime} \left| \Phi \right|^2
 = 0.
\end{equation}
Therefore, should imaginary part of $\omega$ exist, 
$(\kappa^2/\Sigma_0)^{\prime}$ must change the sign somewhere.  This is
Rossby wave instability previously investigated by a number of authors
(Lovelace and Hohlfeld 1978, Lovelace et al. 1999, Li et al. 2000).
Our analysis presented here for non-axisymmetric disk is the
shearing-sheet version of that given by Lovelace and Hohfeldt (1978),
although we have used a different variables from their analysis.
We note that $\kappa^2/\Sigma_0$ is proportional to the background
vortensity, which is $(2\Omega_{\rm p}+U^{\prime})/\Sigma_0$ in the
shearing-sheet approximation.

\clearpage

\clearpage

\clearpage

\clearpage

\clearpage

\clearpage

\clearpage

\clearpage

\clearpage

\clearpage

\clearpage

\clearpage

\clearpage

\clearpage

\clearpage

\clearpage

\begin{table}
\begin{center}
\caption{Mass parameters used in numerical 
 calculations \label{table:parameter}}
\begin{tabular}{cccc}
\tableline\tableline
Model Number & $GM{\rm p}/Hc^2$ 
& Planet Mass\footnote{Assuming $1\mathrm{AU}$ and $H/r_{\rm p}=0.05$.} 
& $r_{\rm H}/H$  \\
\tableline
1 & 0.05 & $1.875 M_{\oplus}$ & 0.26  \\
2 & 0.1 & $3.75 M_{\oplus}$ & 0.32  \\
3 & 0.2 & $7.5 M_{\oplus}$ &  0.41 \\
4 & 0.4 & $15 M_{\oplus}$ &  0.51 \\
5 & 0.6 & $22.5 M_{\oplus}$ & 0.58  \\
\tableline
\end{tabular}
%\tablenotetext{a}{Assuming $1\mathrm{AU}$ and $H/r_{\rm p}=0.05$.}
\end{center}

\end{table}

\end{document}